\documentclass[sigconf,screen,nonacm]{acmart}

\AtBeginDocument{%
  }

\copyrightyear{2024}
\acmYear{2024}
\setcopyright{rightsretained}

\acmConference[LICS '24]{39th Annual ACM/IEEE Symposium on Logic in Computer Science}{July 8--11, 2024}{Tallinn, Estonia}
\acmBooktitle{39th Annual ACM/IEEE Symposium on Logic in Computer Science (LICS '24), July 8--11, 2024, Tallinn, Estonia}\acmDOI{10.1145/3661814.3662126}

\acmISBN{979-8-4007-0660-8/24/07}

\usepackage{subcaption}
\usepackage{tikz-cd}
\usepackage[T1]{fontenc}
\usepackage[utf8]{inputenc}

\newcommand{\dua}{\mathord{\hbox{\makebox[0pt][l]{\raise .6mm
                           \hbox{$\uparrow$}}$\uparrow$}}}
\renewcommand{\mathsf}[1]{#1}
\renewcommand{\mathcal}[1]{#1}

\newcommand{\remove}[1]{}

\newcommand{\ldot}{\, . \,}

\newcommand{\nat}{\mathbb{N}}
\newcommand{\interval}{\mathbb{I}}
\newcommand{\realLine}{\mathbb{R}}
\newcommand{\realDom}{\mathbb{IR}}

\newcommand{\BC}{\textrm{\bf{BC}}}
\newcommand{\D}{\textrm{\bf{D}}}

\newcommand{\PD}{\textrm{\bf{PER}}}

\newcommand{\id}{{\sf id}}

\newcommand{\sle}{\sqsubseteq}

\begin{document}
\title{A Cartesian Closed Category for Random Variables}


\author{Pietro Di Gianantonio}
\affiliation{%
  \institution{Univesity of Udine}
  \city{Udine}
  \country{Italy}
}
\email{pietro.digianantonio@uniud.it}

\author{Abbas Edalat}
\affiliation{%
  \institution{Imperial College London}
  \city{London}
  \country{UK}}
\email{ae@ic.ac.uk}


\begin{abstract}
We present a novel, yet rather simple construction within the traditional framework of Scott domains to provide semantics to probabilistic programming, thus obtaining a solution to a long-standing open problem in this area.  We work with the Scott domain of random variables from a standard and fixed probability space---the unit interval or the Cantor space---to any given Scott domain. The map taking any such random variable to its corresponding probability distribution provides a Scott continuous surjection onto the probabilistic power domain of the underlying Scott domain, which preserving canonical basis elements, establishing a new basic result in classical domain theory. If the underlying Scott domain is effectively given, then this map is also computable. We obtain a Cartesian closed category by enriching the category of Scott domains by a partial equivalence relation to capture the equivalence of random variables on these domains. The constructor of the domain of random variables on this category, with the two standard probability spaces, leads to four basic strong commutative monads, suitable for defining the semantics of probabilistic programming.
\end{abstract}

\maketitle

\section{Introduction}

Probabilistic programming languages (PPLs) have recently attracted considerable interest due to their applications in areas such as machine learning, artificial intelligence, cognitive science, and statistical modelling. These languages provide a powerful framework for specifying complex probabilistic models and performing automated Bayesian inference.

In parallel with these developments, there has been a growing interest in defining a formal (denotational) semantics for these languages. This formalisation is crucial for understanding the theoretical foundations of PPLs and for ensuring the correctness and efficiency of the computations they perform.

The pursuit of a robust semantic foundation for PPLs through domain theory not only enhances our understanding of these languages but also opens new avenues for advancing their capabilities and applications. By grounding probabilistic programming in solid mathematical theory, we can unleash its full potential in various areas of computation.

This endeavour essentially boils down to defining an appropriate probability monad that can be used to represent the result of a probabilistic computation. A probability monad encapsulates the probabilistic aspects of computations, allowing the integration of uncertainty and stochastic behaviour in a rigorous way.

In the context of domain theory, which is our focus, the traditional approach in denotational semantics of programming languages is based on using Scott domains or more generally continuous dcpo's  (directed complete partial orders), which are equipped with the fundamental notion of approximation represented by the way-below relation inherent in these mathematical structures. In analogy with the three main power domains for non-determinism, the use of the probabilistic power domain has been the standard construction for probabilistic computation. This concept, introduced in the seminal works of Saheb-Djahromi (1980)~\cite{Saheb80} and Jones and Plotkin (1989)~\cite{JonesP89}, extends the classical power domain constructions to handle probabilistic computations. The probabilistic power domain allows the modelling of probabilistic choice and uncertainty within a domain-theoretic framework.

This approach, however, soon hit a stumbling block. In fact, a long-standing open problem, known as the Jung-Tix problem, in this area has been to obtain an appropriate category of continuous dcpo's that is a Cartesian closed category (CCC) and closed with respect to the probabilistic power domain constructor: none of the appropriate categories are known to support function spaces~\cite{JungT98}. 

In the absence of a CCC of continuous dcpo's, closed under the probabilistic power domain, several domain-theoretic researchers have been led to abandon classical domain theory and attempt to use more general categories in this area, by relaxing the condition of working with continuous dcpo's, i.e., using a category that contains non-continuous dcpo's as in~\cite{JiaLMZ21} and~\cite{GoubaultJT23}. However, this means giving up the celebrated notion of approximation via the way-below relation in continuous dcpo's, traditionally a staple in domain theory and denotational semantics. Other researchers have embarked on defining or using new mathematical structures, such as quasi-Borel spaces, for developing a model of PPL; see below.

In our work, we propose to work within the simple category of Scott domains, but we circumvent the Jung-Tix problem as follows: instead of describing a computation as a probability distribution, or a continuous valuation, over the space of possible results, we describe a computation as a random variable, from a fixed standard sample space to the space of possible results. The sample space can be considered as the source of randomness used by an otherwise deterministic computation to generate probabilistic behaviour.

Random variables are a key concept in probability theory, having a status as fundamental as probability distributions, and descriptions of probabilistic computation can be naturally expressed in terms of them. They have been employed in denotational semantics of programming languages in various work \cite{Barker16,Mislove16,GoubaultV11,Scott14,HeunenKSY17,Vakar2019,HuotLMS23}.

In fact, our first main result, on which the whole work is built, shows that the probability map from the Scott domain of random variables on a Scott domain that takes any random variable to its associated probability distribution in the probabilistic power domain of the Scott domain is a Scott continuous surjection, which preserves the canonical basis elements of the two domains. This gives a simple many-to-one correspondence between random variables and probability distributions on a Scott domain. It thus gives a new representation of the probabilistic power domain of a given Scott domain by the Scott domain of the random variables on it.

In addition, if a Scott domain $D$ is effectively given~\cite{smyth1977effectively,plotkin1981post}, then its probabilistic power domain and the Scott domain of random variables on $D$ are also effectively given, and the probability map would be computable. It is a distinct feature of our work that the basis of Scott domains can be used to characterize computable random variables and probability distributions on the Scott domains. 

This new representation allows us to develop, for PPLs, a CCC based on Scott domains, thus providing a solution to the long-standing open problem in this area. Our constructions are simply a combination of Scott domains with a partial equivalence relation to capture the equivalence of random variables corresponding to the same probability distribution.  In several other contexts, the concept of using equivalence relation in denotational semantics is quite standard and widespread\cite{AP90,Danielsson2006,sabelfeld2001per}.

We employ both the Cantor space with its standard uniform product measure and the unit interval with its uniform (Lebesgue) measure as our probability spaces, which provide us with four canonical, strong commutative monads for constructing random variables on Scott domains. As in classical probability theory, the same probability distribution can be described by different random variables.  To furnish a coherent use of equivalent random variables, we enrich our domains with partial equivalence relations. More generally, we use a category whose morphisms are equivalence classes of functions, rather than single functions between objects.  This is a main point that distinguishes our approach from mainstream denotational semantics.
Consequently, we allow a multiplicity of possible semantics for the same computation, and, as in the case of a two-category, the commutative diagrams characterising the probabilistic monads hold up to an equivalence relation on functions.

The equivalence relation on random variables needs to be defined also on domains built by the repeated application of the random variable constructor, for example on the domain of random variables constructed on the domain of random variables on real numbers. This is achieved by introducing a new natural topology on domains, called the R-topology, consisting of Scott open sets invariant with respect to the partial equivalence relation.

We show, by various examples, in the last section that we can define random variables corresponding to basic probability distributions on finite dimensional Euclidean spaces, and that functions of random variables such as the Dirichlet distribution can also be expressed in the framework. Since we are using Scott domains, we have a foundation for PPLs based on exact computation for evaluating elementary functions~\cite{DiGianantonio1997a,Esc96,potts1997semantics}. 

The remainder of the paper is organised as follows. In the rest of this section, we review some related work focusing on those using random variables. In Section~\ref{prelim}, we present the basic notions, properties and constructions in domain theory and measure theory we require in this paper. In Section~\ref{rv-cv}, we present four canonical probability spaces, constructed from the Cantor space and the unit interval, used to define the Scott domain of  random variables on a given Scott domain. We show that the probability map which takes a random variable to its associated probability distribution in the probabilistic power domain of the Scott domain preserves canonical basis elements and is an effectively given continuous surjection.  In Section~\ref{monad-sec}, the category of Scott domain is enriched with a partial equivalence relation, which is shown to give a CCC called {\bf PER}. It is then shown that strong and commutative monads of random variables can be constructed on {\bf PER} using the R-topology
. 
In Section~\ref{rv-r}, we present random variables for various probability distributions on finite dimensional Euclidean spaces. 
\subsection{Related works}

In the existing literature, there are a limited number of constructions of a probabilistic monad based on random variables, and notably, all of them use definitions that are significantly different from ours. In~\cite{Barker16}, built from ideas in~\cite{GoubaultV11}, a Random Choice Functor (RC) is proposed, which uses an alternative representation for random variables. In this framework, a random variable is defined as a pair consisting of a domain and a function. This approach leads to a scenario where a single random variable, in our setting, corresponds to several different representations in the RC framework. More critically, the morphisms defining the monad on RC do not preserve the equivalence relation between random variables as defined in our paper. Consequently, these morphisms have no correspondence in our setting. In addition, there is no discussion of the commutativity of the monad in~\cite{Barker16}.
Domains of random variables with a structure quite similar to that of the present paper are defined in~\cite{Mislove16,Scott14,BacciFKMPS18}, but no monad construction was presented there. 

Other works that define the probabilistic monad through the notion of random variables include \cite{HeunenKSY17, Vakar2019, HuotLMS23}. The semantic structures considered in these works are quite different from ours. Quasi-Borel, and $\omega$-quasi-Borel spaces are used in \cite{HeunenKSY17,Vakar2019}.  An ($\omega-$)quasi-Borel space is characterised by specifying the sets of random variables from the sample spaces to the given space; in these spaces, random variables are primitive notions. In contrast, in our work, the random variables are defined in terms of the cpo structure. 
A somewhat similar approach is used in \cite{HuotLMS23}, where $\omega$-PAPs are spaces characterised by specifying sets of piecewise analytic functions. 
The construction of the probabilistic monads in these works is quite different from ours, although parallels can be drawn between our power domain constructions and those in \cite{Vakar2019}. Both approaches have to deal with the fact that equivalence classes of random variables (i.e., sets of random variables defining the same measure) are not necessarily complete under the limits of $\omega$-chains. In \cite{Vakar2019}, the problem is solved by embedding the space of measures in a larger $\omega$-complete space, whereas we address the problem by working only with $\omega$-chains of random variables and avoiding dealing with $\omega$-chains of equivalence classes.

{There is a large literature on denotational models for probabilistic computation.  These works differ substantially from ours, firstly by not using random variables and instead modelling probabilistic computation using probabilistic measures \cite{DanosE11,EhrhardPT18} or continuous distributions \cite{GoubaultJT23,JiaLMZ21}; and secondly by the fact that these works do not use Scott domains but instead a larger class of dcpo's \cite{GoubaultJT23,JiaLMZ21} or other semantic structures \cite{DanosE11,EhrhardPT18}.}

\section{Domain-theoretic preliminaries}\label{prelim}

We first present the elements of domain theory and topology required here; see~\cite{abramsky1995domain} and~\cite{gierz2003continuous} for references to domain theory. 
\subsection{Some basic notions and properties} We denote the interior and the closure of a subset $S$ of a topological space by $S^\circ$ and $\overline{S}$ respectively. The lattice of open sets of a topological space $X$ is denoted by $\tau_{\scriptscriptstyle X}$; it is always clear from the context which topology for $X$ is meant by $\tau_{\scriptscriptstyle X}$. For a map $f:X\to Y$, denote the image of any subset $S\subseteq X$ by $f[S]$. If $I\subseteq \realLine$ is a non-empty real interval, its left and right endpoints are denoted by $I^-$ and $I^+$ respectively; thus if $I$ is compact, $I=[I^-,I^+]$. A {\em dyadic} number is of the form $m/2^n$, for some $m,n\in \nat$.

A {\em directed complete partial order} (dcpo) $D$ is a partial order in which every (non-empty) directed set $A\subseteq D$ has a lub (least upper bound) or supremum $\sup A$. The {\em way-below relation} $\ll$ in a dcpo $(D,\sqsubseteq)$ is defined by $x\ll y$ if whenever there is a directed subset $A\subseteq D$ with $y\sqsubseteq \sup A$, then there exists $a\in A$ with $x\sqsubseteq a$. An element $x\in D$ is {\em compact} if $x\ll x$. A subset $B\subseteq D$ is a {\em basis} if for all $y\in D$ the set $\{x\in B:x\ll y\}$ is directed with lub $y$. By a {\em domain}, we mean a non-empty dcpo with a countable basis. Domains are also called {\em $\omega$-continuous dcpo's}. If the basis of a domain consists of compact elements, then it is called an {\em $\omega$-algebraic dcpo}. In a domain $D$ with basis $B$, we have the interpolation property: the relation $x\ll y$, for $x,y\in D$, implies there exists $z\in B$ with $x\ll z\ll y$. 

A subset $A\subseteq D$ is {\em bounded} if there exists $d\in D$ such that for all $x\in A$ we have $x\sqsubseteq d$. If any bounded subset of $D$ has a lub then $D$ is called {\em bounded complete}. Thus, a bounded complete domain has a bottom element $\bot$, which is the lub of the empty subset. A bounded complete domain $D$ has the property that any non-empty subset $S\subseteq D$ has an infimum or greatest lower bound, denoted $\inf S$. 

The set of non-empty compact intervals of the real line ordered by reverse inclusion and augmented with the whole real line as bottom is the prototype bounded complete domain for real numbers denoted by $\realDom$, in which $I\ll J$ iff $J \subseteq I^\circ$. It has a basis consisting of all intervals with rational endpoints.  For two non-empty compact intervals $I$ and $J$, their infimum $I\sqcap J$ is the convex closure of $I\cup J$. The Scott topology on a domain $D$ with basis $B$ has sub-basic open sets of the form $\dua b:=\{x\in D:b\ll x\}$ for any $b\in B$. 

The lattice $\tau_D$ of Scott open sets of a domain $D$ is continuous. The basic Scott open sets for $\realDom$ are of the form $\{J\in \realDom:J\subseteq I^\circ\}$ for any $I\in \realDom$. The maximal elements of $\realDom$ are the singletons $\{x\}$ for $x\in \realLine$ which we identify with real numbers, i.e., we write $\realLine\subseteq \realDom$, as the mapping $x\mapsto \{x\}$ is a topological embedding when $\realLine$ is equipped with its Euclidean topology and $\realDom$ with its Scott topology. Similarly, $\interval [a,b]$ is the domain of non-empty compact intervals of $[a,b]$ ordered with reverse inclusion. 

If $X$ is any topological space with some open set $O\subseteq X$ and $d\in D$, then the single-step function $d\chi_O:X\to D$, defined by $d\chi_O(x)=d$ if $x\in O$ and $\bot$ otherwise, is a Scott continuous function. The partial order on $D$ induces, by pointwise extension, a partial order on continuous functions of type $X\to D$ with $f\sqsubseteq g$ if $f(x)\sqsubseteq g(x)$ for all $x\in X$.  For any two bounded complete domains $D$ and $E$, the function space $(D\to E)$ consisting of Scott continuous functions from $D$ to $E$ with the extensional order is a bounded complete domain with a basis consisting of lubs of bounded and finite families of single-step functions. We will use the following properties, as presented in the remainder of this subsection, widely in the paper.
\begin{lemma}\label{way-below-func}\cite[Proposition II-4.20(iv)]{gierz2003continuous}
Suppose $X$ is a topological space such that its lattice $\tau_X$ of open sets is continuous and $D$ is a bounded complete domain. If $f:X\to D$ is Scott continuous and $d\chi_O:X\to D$ is a single-step function, then 
\[d\chi_O\ll f \iff O\ll_{\tau_X} f^{-1}(\dua d)\]
\end{lemma}
The next property, which we will use in the construction of the monads as well as in deriving random variables with a given probability distribution on $\realLine$,  is a consequence of the distinct feature of Scott domains as densely injective spaces~\cite{gierz2003continuous}.
\begin{proposition}\label{envelope} \cite[Exercise II-3.19]{gierz2003continuous}
 If $h:A\subseteq Y\to D$ is any map from a dense subset $A$ of $Y$ into a bounded complete domain $D$, then its \emph{envelope} 
 \[h^\star:Y\to D\]
 given by $h^\star(x)=\sup\{\inf h[O]: x\in O \mbox{ open}\}$ is a continuous map with $h^\star(x)\sqsubseteq h(x)$, and in addition $h^\star(x)=h(x)$ if $h$ is continuous at $x\in A$. Moreover, $h^\star$ is the greatest continuous function $p:Y\to D$ with $p(x)\sqsubseteq h(x)$ for all $x\in Y$.
\end{proposition}
Since $\realLine\subseteq \realDom$ is dense, any continuous map $f:\realLine\to \realLine\subseteq \realDom$, considered as a continuous map $f:\realLine\to \realDom$, has a maximal extension $f^\star:\realDom\to\realDom$ given by $f^\star(x)=f[x]$, i.e., the pointwise extension of $f$ to compact intervals. 
\begin{lemma}\label{map-frame-map}
 Suppose $X$ is a topological space and $D$ a dcpo. If $r_i:X\to D$ is a directed set of Scott continuous functions with $r=\sup_{i\in I} r_i$ and $O\subseteq D$ is Scott open then $r^{-1}(O)=\bigcup_{i\in I}r_i^{-1}(O)$.  
\end{lemma}
\begin{proof}
 From $r_i\sqsubseteq r$ we obtain $r_i^{-1}(O)\subseteq r^{-1}(O)$ for $i\in I$. Thus, we have $r^{-1}(O)\supseteq\bigcup_{i\in I}r_i^{-1}(O)$. If $x\in r^{-1}(O)$, then $\sup_{i\in I}r_i(x)=r(x)\in O$ and since $O$ is inaccessible by any directed set, there exists $i\in I$ such that $r_i(x)\in O$, and the result follows. 
\end{proof}

A {\em crescent} in a topological space is defined to be the intersection of an open and a closed set. Let $\partial C$ denote the boundary of a subset $C\subseteq X$ of a topological space $X$.
 \begin{proposition}\label{dis-cre}
 
 Suppose $f=\sup_{i\in I} d_i\chi_{O_i}:X\to D$ is a step function from a topological space $X$ to a bounded complete domain $D$. Then we have $f=\sup_{j\in J} c_j\chi_{C_j}$ where $c_j$ for $j\in J$ are the distinct values of $f$ and $C_j$ for $j\in J$ are disjoint crescents, generated from $O_j$ with $j\in J$ by the two operations of taking finite intersections and set difference.  Moreover, if $x\in \partial C_k$ for some $k\in J$, then, by Scott continuity, we have $f(x)=\inf\{ c_j:x\in \partial C_j\}$.
 \end{proposition}

\subsection{Normalised probabilistic power domain} 

 Recall
from~\cite{birkhoff1940lattice,Saheb80,lawson1982valuations,JonesP89} that a {\em valuation} on a topological
space $Y$ is a map
$\nu:\tau_Y\to [0,1]$ which satisfies:
\begin{itemize}
\item[(i)] $\nu(a)+\nu(b)=\nu(a\cup b)+\nu(a\cap b)$
\item[(ii)] $\nu(\emptyset)=0$
\item[(iii)] $a\subseteq b\to \nu(a)\leq \nu(b)$
\end{itemize}

A {\em continuous} valuation~\cite{lawson1982valuations,JonesP89} is a valuation such that whenever
$A\subseteq\tau_{Y}$ is a directed set (wrt $\subseteq$) of open sets
of $Y$, then
\[\nu\left(\bigcup_{O\in A}O\right)=\mbox{sup}\,_{O\in A}\nu(O).\]

For any $b\in Y$, the {\em point valuation} based at $b$ is the
valuation $\delta_b:\tau_{Y}\to [0,1]$ defined by
\[\delta_b(O)=\left\{\begin{array}{cl}
1&\mbox{if $b\in  O$}\\
0&\mbox{otherwise}.\end{array}\right.\]
Any finite linear combination
$\sum_{i=1}^{n}r_i\delta_{b_i}$
of point valuations $\delta_{b_i}$ with constant coefficients
$r_i\in [0,\infty)$, ($1\leq i\leq n$) is a continuous valuation on
$Y$, called a {\em simple valuation}.

The {\em probabilistic power domain}, $P_0Y$, of a topological space $Y$ consists of
the set of continuous valuations $\nu$ on $Y$ with $\nu(Y)\leq 1$
and is ordered as follows:
\[\mu\sle\nu\mbox{ iff for all open sets $O$ of $Y$, }\mu(O)\leq \nu(O).\]
The partial order $(P_0Y,\sle)$ is a dcpo with bottom in which  the lub of a directed set
$\langle\mu_i\rangle_{i\in I}$ is given by $\sup_i\mu_i=\nu$, where
for $O\in \tau_{Y}$:
\[\nu(O)=\mbox{sup}\,_{i\in I}\mu_i(O).\]

\subsubsection{Normalized continuous valuations}
Let $\D$ be the category of countably based domains, also known as $\omega$-continuous domains.
 We will work with normalised continuous valuations of a domain $D\in \D$, i.e., those with unit mass on the whole space $D$. These will correspond to probability distributions on $D$. Consider the {\em normalised} probabilistic power domain $PD$ of $D\in \D$, consisting of normalised continuous valuations with pointwise order. Like $P_0D$ as shown in~\cite{JonesP89},  $PD$ is an $\omega$-continuous dcpo, that is an object in $\D$, with a countable basis consisting of normalized valuations given by a finite sum of pointwise valuations with rational coefficient~\cite{edalat1995domain}.

 Unless otherwise stated, all continuous valuations in this paper are normalised continuous valuations and by the probabilistic power domain $PD$ of domain $D$, we always mean the normalised probabilistic power domain. 

The splitting lemma for normalised valuations~\cite{edalat1995domain}, which is similar to the splitting lemma for valuations~\cite{JonesP89}, states: If \[\alpha=\sum_{1\leq i\leq m}p_i\delta(c_i),\quad\beta=\sum_{1\leq j\leq n}q_j\delta(d_j),\]are two normalised valuations on a continuous dcpo $D$ then $\alpha\sqsubseteq\beta$ iff there exist $t_{ij}\in [0,1]$ for $1\leq i\leq m$, $1\leq j\leq n$ such that
\begin{itemize}
\item $\sum_{i=1}^m t_{ij}=q_j$ for each $j=1, \ldots, n$.
\item $\sum_{j=1}^n t_{ij}=p_i$ for each $i=1, \ldots, m$.
\item $t_{ij}>0\,\Rightarrow\,c_i\sqsubseteq d_j$. 
\end{itemize}
We say $t:=(t_{ij})_{i\in I, j\in J}$ is a {\em flow} from $\alpha$ to $\beta$ witnessing, $\alpha\sle \beta$ and we write $t:\alpha\to \beta$.

We also have a splitting lemma for the way-below relation on normalised valuations. Suppose $\alpha=\sum_{1\leq i\leq m}p_i\delta(c_i)$ and $\beta=\sum_{1\leq j\leq n}q_j\delta(d_j)$ are normalised valuations on a continuous dcpo. 
\begin{proposition}\label{sp-way}\cite[Proposition 3.5]{edalat1995domain} We have $\alpha\ll\beta$ if and only if there exist $t_{ij}\in [0,1]$ for $1\leq i\leq m$ and $1\leq j\leq n$ such that 
\begin{itemize}
\item $c_{i_0}=\bot$ for some $i_0$ with $1\leq i_0\leq  m$ and for all $j$ with $1\leq j\leq m$, we have $t_{i_0j}>0$,
\item $\sum_{i=1}^m t_{ij}=q_j$ for each $j=1, \ldots, n$,
\item $\sum_{j=1}^n t_{ij}=p_i$ for each $i=1, \ldots, m$.\
\item $t_{ij}>0\,\Rightarrow c_i\ll d_j$. 
\end{itemize}
\end{proposition}

For a basis $B_D$ for $D$, fix a canonical basis $B_{PD}$ of normalised simple valuations for $PD$ consisting of \emph{normalised simple valuations} 
\begin{equation}\label{can-base}\sum_{0\leq i\leq m}p_i\delta(c_i)\end{equation}
with $c_i\in B_D$, $p_i$ dyadic, for $1\leq i\leq m$ with $\sum_{i=0}^mp_i=1$, and $c_0=\bot$ with $p_0>0$. By Proposition~\ref{sp-way}, this basis has the property that the set given by $\{\dua \sigma:\sigma\in B_{PD}\}$ is a basis for $\tau_{PD}$. 

\begin{proposition}\label{simp-val-way}
    Suppose $D$ is a continuous dcpo with a simple valuation  $\sigma=\sum_{i\in I} p_i\delta(d_i)\in PD$  and  a continuous valuation $\alpha\in PD$. Then $\sigma\ll \alpha$ iff there exists $i_0\in I$ with $d_{i_0}=\bot$ such that for all $J\subseteq I\setminus \{i_0\}$ we have: 
    \begin{equation}\label{simp-way-below-eq}\sum_{j\in J}p_j<\alpha\left(\bigcup_{j\in J} \dua d_j\right)\end{equation} 
\end{proposition}
\begin{proof}
    Suppose $\sigma\ll \alpha$. Take any simple valuation $\sigma'$ with $\sigma\ll \sigma'\ll \alpha$. Using Proposition~\ref{sp-way}, since $\sigma'(U) \leq \alpha(U)$ for any open set $U\subseteq D$, we obtain: 
    \[\sum_{j\in J}p_j<\sigma'\left(\bigcup_{j\in J} \dua d_j\right) \leq \alpha\left(\bigcup_{j\in J} \dua d_j\right)\]
  
    Next suppose Equation~(\ref{simp-way-below-eq}) holds.  Let $\sigma_0=\sum_{i\in I\setminus\{i_0\}} p_i\delta(d_i)\in P_0D$, where $P_0D$ is the probabilistic power domain of continuous valuation whose total mass is bounded by 1.  By~\cite[p. 46]{kirch1993bereiche}, we have $\sigma_0\ll \alpha$ in $P_0D$ and it then follows by~\cite[Corollary 3.3]{edalat1995domain} that $\sigma\ll \alpha$ in $PD$ as required. 
\end{proof}
From Proposition~\ref{sp-way}, it follows that $\delta(d) \ll \sum_{1\leq j\leq n}q_j\delta(d_j)$ iff $d=\bot$. Thus, the simplest non-trivial simple valuation way below another simple valuation takes the form $p\delta (d) +(1-p)\delta(\bot)\ll \sum_{i\in I}q_i\delta(d_i)$ for $p<1$ and $d\neq \bot$, and we have the simple property: 
\begin{corollary}\label{simplest-simple-way-below}
If $0\leq p<1$ and $I$ is a finite set, then:
\[p\delta(d)+(1-p)\delta(\bot)\ll \sum_{i\in I}q_i\delta(d_i)\;\iff \;p<\sum_{d\ll d_i} q_i\]
\end{corollary}

\subsection{Measure theory and domain theory}\label{m-d} Recall, say from~\cite{athreya2006measure}, that a measurable space on a set $X$ is given by a $\sigma$-algebra $S_X$ subsets of $X$, i.e., a non-empty family of subsets of $X$ closed under the operations of taking countable unions, countable intersections and complementation. Elements of $S_X$ are called measurable sets. For a topological space $X$, the collection of all Borel sets on X forms a $\sigma$-algebra, known as the $\sigma$-Borel algebra: it is the smallest $\sigma$-algebra containing all open sets (or, equivalently, all closed sets). A map $f: (X,S_X)\to (Y,S_Y)$ of two measurable spaces is measurable, if $f^{-1}(C) \in S_X$ for any $C\in S_Y$. Any continuous function of topological spaces is measurable with respect to the Borel algebras of the two spaces. 

A probability measure on a measure space $(X,S_X)$ is a map $\nu:S_X\to [0,1]$ with $\nu(\emptyset)=0$, $\nu(X)=1$ such that $\nu(\bigcup_{i\in \nat}\nu(E_i))=\sum_{i\in \nat}\nu(E_i)$ for any countable disjoint collection of measurable sets $E_i$ for $i\in \nat$. A {\em probability space} $(\Omega,S_\Omega,\nu)$ is given by a probability measure $\nu$ on a measure space $(\Omega,S_\Omega)$, where $\Omega$ is called the {\em sample space} and measurable sets are called {\em events}. A subset $A\subseteq \Omega$ is a {\em null set} if $A\subseteq B\in S_\Omega$ with $\nu(B)=0$. 

Given the probability space $(\Omega,S_\Omega,\nu)$, a random variable on a measurable space $(Y,S_Y)$ is a measurable map $r: (\Omega,S_\Omega,\nu)\to (Y,S_Y)$. The probability of $r\in C$ for $C\in S_Y$ is given by $\operatorname{Pr}(r\in C)= \nu(r^{-1}(C))$. The probability measure $\nu\circ r^{-1}$ is the probability distribution or {\em push-forward measure} induced by $r$. Two random variables $r_1$ and $r_2$ are {\em independent} if $\operatorname{Pr}(r_1\in C_1,r_2\in C_2)= \operatorname{Pr}(r_1\in C_1)\operatorname{Pr}(r_2\in C_2)$ for all $C_1,C_2\in S_Y$. Two independent and identically distributed random variables are denoted as i.i.d. 

For two measurable functions $f,g: (\Omega,S_\Omega,\nu)\to (Y,S_Y)$, we say $f=g$ almost everywhere, written as a.e., if the set of points on which they are not equal is a null set.  A measurable map $f:(\Omega,S_\Omega,\nu_\Omega)\to (\Omega',S_{\Omega'},\nu_{\Omega'})$ of two probability spaces is measure-preserving if $\nu_X(f^{-1}(C))=\nu_Y(C)$ for $C\in S_Y$. The theory of Lebesgue integration is built on measure spaces; see~\cite{walter1987real}.

For a countably based continuous dcpo, every probability measure extends uniquely to a continuous valuation, as is easily checked. Conversely, for such spaces, every continuous valuation extends uniquely to a probability measure~\cite{alvarez2000extension}. Similarly, every continuous valuation on a countably based locally compact Hausdorff space extends uniquely to a measure on the space~\cite{lawson1982valuations}. Moreover, any probability measure $\nu$ on a countably based locally compact Hausdorff space $\Omega$ is regular, i.e., $\nu(S)=\inf \{\nu(O):S\subseteq O\in \tau_{\scriptscriptstyle \Omega}\}$~\cite{walter1987real}. 

By these results, without imposing any limitations, we can work with open sets, as events, and normalised continuous valuations rather than general Borel or measurable sets, and probability measures~\cite{halmos2013measure,lawson1982valuations,Edalat1995}. This corresponds to the notion of an open set as a semi-decidable or observable predicate as formulated in~\cite{smyth1983power} and~\cite{abramsky1991domain,jung2013continuous}, the underlying basis of observational logic~\cite{vickers1989topology}. In the domain-theoretic framework of observational logic, as we adopt in this work, we use open sets instead of measurable sets and continuous functions, instead of measurable functions, as random variables for probabilistic computation.  

\subsection{Effectively given domains}\label{effect}
A countably based domain $D\in\D$ can be equipped with an effective structure~\cite{smyth1977effectively,plotkin1981post}. We will provide a concise account of this as presented in~\cite{edalat2002foundation}. For bounded complete domains, this formulation is equivalent to that in~\cite{plotkin1981post} except that the latter is formulated for algebraic, rather than continuous, Scott domains. We say $D$ is {\em effectively given} with respect to an enumeration $b:{\mathbb N}\to B$ of the countable basis $B\subseteq D$ if the set $\{\langle m,n\rangle: b_m\ll b_n\}$ is decidable, where $\langle-,-\rangle:{\mathbb N}\times {\mathbb N}\to {\mathbb N}$ is the standard pairing function, i.e., the isomorphism $\langle m,n\rangle=(m+n)(m+n+1)/2 +m$. If $D$ has a least element $\bot$, we assume $b_0=\bot$. If $D$ is, additionally, bounded complete, we further assume that the set $\{\langle \langle m,n\rangle,p\rangle:b_m\sqcup b_n=b_p\}$ is decidable. We say $x\in D$ is {\em computable} if the set $\{n:b_n\ll x\}$ is r.e., which is equivalent to the existence of a total recursive function $g:\nat\to \nat$ such that $(b_{g(n)})_{n\in \nat}$ is an increasing sequence with $x=\sup_{n\in\nat}b_{g(n)}$. \remove{A sequence $(x_n)_{n\in \nat}$ is said to be {\em computable} if there exists a total recursive function $h$ such that $x_n=x_{h(n)}$ for $n\in \nat$.} A continuous map $f:D\to E$ of effectively given domains $D$, with basis $\{a_0,a_1,\ldots\}$, and $E$, with basis $\{b_0,b_1\ldots\}$, is {\em computable} if $\{\langle m,n\rangle:f(a_m)\ll b_n\}$ is r.e. 
\subsection{A domain model for the Hilbert curve}\label{hilbert-sec} 
\begin{figure}[h]
  \centering
  \begin{subfigure}{.14\textwidth}
    \centering
    \includegraphics[width=\linewidth]{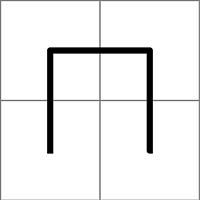}
    \caption*{$i = 1$}
    \label{fig:sub1}
  \end{subfigure} \hspace{1ex}%
  \begin{subfigure}{.14\textwidth}
    \centering
    \includegraphics[width=\linewidth]{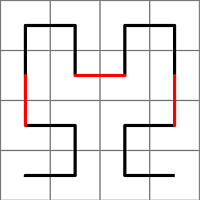}
    \caption*{$i = 2$}
    \label{fig:sub2}
  \end{subfigure} \hspace{1ex} %
  \begin{subfigure}{.14\textwidth}
    \centering
    \includegraphics[width=\linewidth]{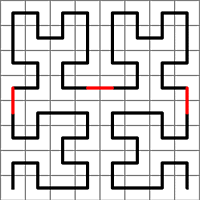}
    \caption*{$i = 3$}
    \label{fig:sub3}
  \end{subfigure}
  \caption{Partition of the unit square by the grid $G_i$ having $4^i$ sub-squares with digit strings of length $i \in \{1,2,3\}$. The connected piecewise linear curve in the interior of the square for $i=1$ is mapped by $H_j$ for $j=1,2,3,4$ to the four black piecewise linear curves in the square for $i=2$. Similarly, the connected piecewise linear curve in the interior of the square for $i=2$ is mapped by $H_j$ for $j=1,2,3,4$ to the four black piecewise linear curves in the square for $i=3$.}
  \label{hilbert-curve}
\end{figure}

Two out of the four canonical monads presented in this paper are based on the well-known Hilbert's space-filling curve, which has been widely used in different branches of computer science~\cite{bader2012space}. It provides a continuous, measure-preserving surjective map of the unit interval to the unit square that is a bijection on a set of full measure. We use a simple representation of this curve by an iterated function system (IFS) presented first in \cite{sagan1992geometrization} to construct a domain-theoretic model of the curve~\cite{Edalat1995}; see also~\cite{sagan2012space}. We take the quaternary representation of real numbers in $[0,1]$, so that $\omega\in [0,1]$ is represented by \[0._4\omega_0\omega_1\omega_2\ldots=\sum_{i\in \nat} \omega_i/4^i,\] where $\omega_i\in \{0,1,2,3\}$. This representation is unique if we stipulate, as usual, that no non-zero number can have a representation ending with infinite sequences of $0$'s.  We obtain four affine maps $h_d:[0,1]\to [0,1]$ constructed by the four digits $d=0,1,2,3$ given by \[h_d(0._4\omega_0\omega_1\omega_2\ldots)=0._4d\omega_0\omega_1\omega_2\ldots+d/4\]
Then the unit interval $[0,1]$ is covered by the four subintervals 
\[[0,1/4],[1/4,1/2],[1/2,3/4],[3/4,1]\]of length $1/4$ given by $h_d[0,1]$ with $d=0,1,2,3$. This idea can be extended to the square $[0,1]^2$, where we also include rotation.
Consider the four affine maps of the unit square $H_i:[0,1]^2\to [0,1]^2$, where $i=0,1,3$, given by

\[{H_0\begin{pmatrix}x\\y\end{pmatrix}=\frac{1}{2}\begin{pmatrix}0&1\\1&0
\end{pmatrix}
\begin{pmatrix}x\\y\end{pmatrix},\,H_1\begin{pmatrix}x\\y\end{pmatrix}=\frac{1}{2}\begin{pmatrix}1&0\\0&1
    \end{pmatrix}\begin{pmatrix}x\\y\end{pmatrix}+\frac{1}{2}\begin{pmatrix}0\\1\end{pmatrix}}\]
\[{H_2\begin{pmatrix}x\\y\end{pmatrix}=\frac{1}{2}\begin{pmatrix}1&0\\0&1
    \end{pmatrix}\begin{pmatrix}x\\y\end{pmatrix}+\frac{1}{2}\begin{pmatrix}1\\1\end{pmatrix}}\]
    \[{H_3\begin{pmatrix}x\\y\end{pmatrix}=\frac{-1}{2}\begin{pmatrix}0&1\\1&0
    \end{pmatrix}\begin{pmatrix}x\\y\end{pmatrix}+\frac{1}{2}\begin{pmatrix}2\\1\end{pmatrix}.}\]
    Each $H_i$ takes the unit square bijectively to a sub-square as follows:\[H_0[0,1]^2=[0,.5]^2, \quad H_1[0,1]^2=[0,.5]\times [.5,1],\]\[ H_2[0,1]^2=[.5,1]\times[.5,1],\quad H_3[0,1]^2=[.5,1]\times [0,.5].\]
    As in~\cite{Edalat1995}, we obtain a Scott continuous map  $H:\interval [0,1]^2\to \interval [0,1]^2$, where $\interval[0,1]^2$ is the bounded complete domain of subsquares of the unit square, partially ordered by reverse inclusion, defined by \[H(S)=\bigcup_{0\leq i\leq 3}H_i[S].\] It induces an IFS tree~\cite{edalat1996power}, such that the infinite sequence in $\omega=0._4\omega_0\omega_1\omega_2\ldots=\sum_{i\in \nat} \omega_i/4^i$ gives the infinite branch: 
    \[H_{\omega_0}[0,1]^2\supset H_{\omega_0}H_{\omega_1}[0,1]^2\supset \cdots \supset H_{\omega_0}H_{\omega_1}\ldots H_{\omega_{i-1}}[0,1]^2\supset\cdots\]
    with $H_{\omega_0}H_{\omega_1}\ldots H_{\omega_{i-1}}[0,1]^2$  a subsquare of dimensions $2^{-i}\times 2^{-i}$ for each $i\in \nat$. The collection of all the $4^i$ sub-squares for fixed $i\in \nat$ gives a grid $G_i$ of points that lie on $2(2^i+1)$ vertical and horizontal line segments of unit length each. Figure~\ref{hilbert-curve} shows the grid $G_3$ partitioning $[0,1]^2$ into $4^i$ sub-squares by strings of digits $0,1,2,3$ of length $i$, with $1 \leq i \leq 3$.
    

   We can now define the map $h:[0,1]\to [0,1]^2$ by \[h(\omega)=\bigcap_{i\in \nat} H_{\omega_0}H_{\omega_1}\ldots H_{\omega_i}[0,1]^2.\] This map is well-defined as it sends the different quaternary representations of the same number to a single point~\cite[p. 16]{sagan2012space}. Then $h$ is a continuous measure-preserving surjection, which is 1-1 except for points in $[0,1]$ that are mapped to a point  in $G_i$ for some $i\in \nat$, where the map is two-to-one or four-to-one. The set $G=\bigcup_{i\in\nat}G_i$ is a null set, and so is the set $S\subseteq [0,1]$ of points that are mapped to $\mathcal{G}$. It follows that the measure-preserving map $h$ restricts to a bijection between $[0,1]\setminus S$ and $[0,1]^2\setminus G$ and has thus a measure-preserving inverse $g:[0,1]^2\setminus G\to [0,1]\setminus S$. 

\section{Random variables and  valuations}\label{rv-cv}

\subsection{Sample and probability spaces}

Let $\Sigma=\{0,1\}$ be the two-point space with its discrete topology. The uniform measure on $\Sigma$ induces the product measure on $\Sigma^\nat$.
\begin{definition}\label{sample-sp}
    Our probability space $\Omega$  is one of the following.  
\begin{itemize}
    \item[(i)] The Cantor space $\Sigma^{\nat}$ of the infinite sequences $\omega=\omega_0\omega_1\cdots$ over bits $0$ and $1$ with the product topology $\tau_{\Sigma^{\nat}}$ and product measure $\nu$ induced on $\Sigma^\nat$ by the uniform distribution on $\{0,1\}$.

    \item[(ii)] The subspace $\Sigma_0^\nat\subseteq \Sigma^\nat$ consisting of all infinite sequences not ending with an infinite sequence of $0$, equipped with the subspace topology and the product measure as in (i). 
    \item[(iii)] The unit interval $[0,1]$ with its Euclidean topology $\tau_{ [0,1]}$ and Lebesgue measure $\nu$.
    \item[(iv)] The open interval $(0,1)\subseteq [0,1]$ with its Euclidean topology and the Lebesgue measure $\nu$.
\end{itemize}\end{definition}
All four probability spaces in Definition~\ref{sample-sp} are Hausdorff and countably based. The two probability spaces $\Sigma^\nat$ and $[0,1]$ are compact, whereas $\Sigma_0^\nat$ and $(0,1)$ are only locally compact. It follows that the probability measures $\nu$ defined on all four spaces are unique extensions of the continuous valuation induced by them~\cite{lawson1982valuations}. Additionally, the probability measure $\nu$ on any of the four spaces $\Omega$ is regular and thus, without any limitations, we can work with open sets rather than arbitrary measurable sets, as we discussed in Subsection~\ref{m-d}.  In practice, the sample spaces $\Sigma^\nat$ and $\Sigma_0^\nat$ would have similar denotational and operational semantics; likewise for $[0,1]$ and $(0,1)$. However, theoretically, as we shall see, the sample spaces $\Sigma_0^\nat$ and $(0,1)$, which have no non-empty compact-open subsets satisfy stronger topological properties. We let $\Omega_0$ denote either the probability space $\Sigma_0^\nat$ or $(0,1)$, whereas $\Omega$ stands for any of four probability spaces. Any element of the four sample spaces is denoted by $\omega\in \Omega$. 

 We next fix a basis of open sets for $\Omega$. Recall that if $\Omega =\Sigma^{\nat}$ or $\Omega =\Sigma^{\nat}_0$, a {\em cylinder set} of length $n\in\nat$ is given by \[[x_0\ldots x_{n-1}]:=\{\omega \in \Omega:\omega_i=x_i,\quad 0\leq i\leq n-1\}\]
 with $x_i\in \Sigma$ for $0\leq i\leq n-1$, with the convention that $n=0$ represents the empty cylinder, i.e., the empty set.
 \begin{definition}\label{basic-open} If $\Omega =\Sigma^{\nat}$ or $\Omega =\Sigma^{\nat}_0$ a {\em basic open set} is a finite union of cylinder sets. If $\Omega =[0,1]$ or $\Omega =(0,1)$, a {\em basic open set} is defined to be a finite union of open intervals with dyadic endpoints. 
 
 \end{definition}
 
We note that for $\Omega =\Sigma^\nat$, the basic open sets are compact, but for $\Omega =\Sigma^\nat_0$ they are non-compact. In fact, $\tau_{\Sigma^\nat}$ is an algebraic lattice, in which the compact elements are the finite unions of cylinder sets; whereas $\tau_{\Sigma_0^\nat}$ is a continuous lattice in which only the empty set is compact. For $\Omega =[0,1]$, the basic open set $[0,1]$ is compact, but for $\Omega =(0,1)$, no basic open set, other than the empty set, is compact. 

By using $\nu$ both for the product measure on $\Sigma^\nat$ or $\Sigma_0^\nat$ and for the Lebesgue measure on $[0,1]$ or $(0,1)$ and by referring to basic open sets of $\Omega$ or $\Omega_0$, we can uniformly state and uniformly prove various results about the four probability spaces. We note that since $\tau_{\scriptscriptstyle \Omega}$, for all cases (i)-(iv), is a countably based continuous lattice, the map $\nu: \tau_{\scriptscriptstyle \Omega} \to \realLine$ is in fact a continuous valuation, i.e., it preserves the lubs of directed sets of open sets.

\subsection{Random Variables and the probability Map}
Let ${\BC}$ be the category of countably based bounded complete domains, also known as (continuous) Scott domains.

\remove{In view of Proposition~\ref{dis-cre} and Definition~\ref{basic-open}, we define a {\em basic crescent} to be a crescent generated from basic open sets by intersection and set difference.}   

Let $D\in \BC$. If $r:\Omega \to D$ is any Scott continuous function and $\nu$ the probability measure on $\Omega$, then $r$ is a random variable on $D$; the push-forward measure $\nu\circ r^{-1}$ on $D$ induced by $r$ restricts to a normalised continuous valuation on $D$. If \begin{equation}\label{simpl-random}
  r=\sup_{1\leq i\leq n}d_i\chi_{O_i},  
\end{equation} is a step function built from basic open sets $O_i$, then $r$ 
 is a random variable with a finite number of values, which we call a {\em simple random variable}. Unless otherwise stated, we assume $d_i$'s are distinct values of $r$. 
Then $\Omega$ is the disjoint union of a finite number of basic crescents in each of which $r$ takes a distinct value, possibly including $\bot$. We denote the collection of simple random variables by $B_{(\Omega\to D)}$, which gives a basis for the function space $(\Omega\to D)$. 
 
\begin{definition}
    The {\em probability map} $T:(\Omega\to D)\to PD$ is defined by $T(r)=\nu\circ r^{-1}$.
Two random variables $r,s\in (\Omega\to D)$ are {\em equivalent}, written $r\sim s$, if $\nu\circ r^{-1}=\nu\circ s^{-1}$ in $PD$.
\end{definition}
Thus, we have $r\sim s \iff T(r)=T(s)$.  There are some simple cases, for which two random variables are equivalent.  If $r=s$ a.e., then clearly $r\sim s$. If $r=\sup_{i\in I}d_i\chi_{O_i}$ is a simple random variable, then $r\sim s$ iff $s=\sup_{i\in I}d_i\chi_{O'_i}$ with $\nu(O_i)=\nu(O_i')$. 
  More generally, for any measure-preserving homeomorphism $t:\Omega \to \Omega$ and any random variable $r:\Omega \to D$, we have $r\sim r\circ t$. 
\begin{lemma}\label{step-simple}
If $r:\Omega \to D$ is a simple random variable in the form $r=\sup_{j\in J}d_j\chi_{C_j}$, where $d_j$'s are distinct for $j\in J$ and the basic crescents $C_j$ are disjoint for $j\in J$, then $\nu(C_j)=\nu(C_j^\circ)$  for $j\in J$ and $T(r)$ is a simple valuation given by
\[T(r)=\sum_{j\in J} \mu(C_j)\delta(d_j).\]
\end{lemma}
\begin{lemma}\label{eq-uniformity}
 Consider any two simple valuations \[\alpha_1=\sum_{i\in I}p_i\delta(c_i)\qquad \alpha_2=\sum_{j\in J}q_j\delta(d_j)\]in $(\Omega\to D)$ with $\alpha_1\sqsubseteq \alpha_2$ and dyadic coefficients $p_i$, $q_i$.
\begin{itemize}
\item[(i)] For any simple random variable $r_1\in (\Omega\to D)$ with $T(r_1)=\alpha_1$, there exists a simple random variable $r_2\in (\Omega\to D)$ with $r_1\sqsubseteq r_2$ and $T(r_2)=\alpha_2$. 
\item[(ii)] For any simple random variable $r_2\in (\Omega\to D)$ with $T(r_2)=\alpha_2$, there exists a simple random variable $r_1\in (\Omega\to D)$ with $r_1\sqsubseteq r_2$ and $T(r_1)=\alpha_1$. In addition, a flow  $(t_{ij})_{i\in I,j\in J}$ witnessing $\alpha_1\sle \alpha_2$ is given,  for $i\in I, j\in J$, by\[t_{ij}=\nu\{\omega \in \Omega: r_1(\omega)=c_i\;\&\; r_2(\omega)=d_j\}.\]
\end{itemize}  
\end{lemma}
\begin{proof}
(i) Let $r_1$ be a step function with $T(r_1)=\alpha_1$.  We can assume $r_1=\sup_{i\in I} c_i\chi_{O_i}$ where $O_i$ are disjoint basic open sets. Since $\alpha_1\sqsubseteq \alpha_2$, by 
the splitting lemma, there exist $t_{ij}$ for $i\in I$ and $j\in J$ such that $p_i=\sum_{j\in J} t_{ij}$ and $t_{ij}>0$ implies $c_i\sqsubseteq d_j$. Since $t_{ij}$ can be obtained from $p_i$ and $q_j$ for $i\in I$ and $j\in J$ from a simple linear system of equations with all coefficients of $t_{ij}$ equal to one, it follows that there exists a solution where $t_{ij}$ is a dyadic number for $i\in I$ and $j\in J$. (In fact, if we apply Gaussian elimination method to this system, we can find a solution by simply swapping or subtracting rows without any need to multiply or divide any row by any number, and therefore the solution for each $t_{ij}$ is obtained simply by adding or subtracting the dyadic numbers $p_i$ for $i\in I$ and $q_j$ for $j\in J$.)  Because $\nu(O_i)=p_i$, it follows that there exist disjoint basic open sets $O_{ij}\subseteq O_i$ for $j\in J_i\subseteq J$ with $O_i=\bigcup_{j\in J_i} O_{ij}$ and $\mu(O_{ij})=t_{ij}>0$. Put $r'_2=\sup_{i\in I,j\in J_i} \{d_j\chi_{O_{ij}}: c_i\sqsubseteq d_j\}$. To obtain a simple random variable that is above $r_1$ on the boundary points of the basic open sets in $r_1$, we let $r_2:=r_2'\sqcup r_1$. Then, $r_1\sqsubseteq r_2$ with $T(r_2)=\alpha_2$.
    
(ii) We can assume $r_2=\sup_{j\in J}d_j\chi_{O_j}$ where $O_j$'s are disjoint open sets with $\nu(O_j)=q_j$ for $j\in J$. Using the splitting lemma for the relation $\alpha_1\sqsubseteq \alpha_2$,  for each $j\in J$, we have $q_j=\sum_{t_{ij}>0} t_{ij}$. Thus, there exists a disjoint partition of the basic open set $O_j=\bigcup\{O_{ij}: t_{ij}>0\}$ into basic open sets $O_{ij}$ with $\nu(O_{ij})=t_{ij}$. Let $O_i=\bigcup_{j\in J} O_{ij}$. Then, \[\nu(O_i)=\sum_{j\in J}t_{ij}=p_i,\] for $i\in I$. Put $r_1=\sup_{i\in I}c_i\chi_{O_i} $, which satisfies $T(r_1)=\alpha_1$, $r_1\sqsubseteq r_2$ and for $i\in I, j\in J$,
    \[t_{ij}=\nu\{\omega \in \Omega: r_1(\omega)=c_i\;\&\; r_2(\omega)=d_j\}.\] 
\end{proof}
 
\remove{
\begin{proposition}\label{eq-arbit-less}
    Suppose we have a family of pairs of equivalent simple random variables $r_i\sim s_i$ for $i\in I$, where $I$ is any finite or infinite set. If there exists a simple random variable $r_0\sqsubseteq r_i$ for all $i\in I$, then there exists a simple random variable $s_0$ with $s_0\sim r_0$ and $s_0\sqsubseteq s_i$ for all $i\in I$.
\end{proposition}}


The two parts (i) and (ii) of Lemma~\ref{eq-uniformity} have a counterpart for $\ll$ where it is assumed that $\alpha_1\ll \alpha_2$ and the simple random variables satisfy: $r_1\ll r_2$. The proofs are similar. We now show a generalisation of these two results in Proposition~\ref{eq-unif-conv} below for item (i) and Proposition~\ref{way-below-random} for item (ii). 
\begin{proposition}\label{eq-unif-conv}
  Suppose $r:\Omega \to D$ is a random variable with a simple random variable $r_0\ll r$. If $s_0\sim r_0$, then there exists a random variable $s:\Omega \to D$ with $s_0\ll s$ and $r\sim s$.
  
\end{proposition}
{\begin{proof}
  Let $r_i\ll r_{i+1}$ for $i\in\nat $ be an increasing chain with $r=\sup_{i\in \nat} r_i$. By Proposition~\ref{eq-uniformity} applied recursively, we can find simple random variables $s_i$ with $s_{i+1}\sim r_{i+1} $ for $i\in \nat$ such that $s_i\ll s_{i+1}$ is for $i\in \nat$. Let $s=\sup_{i\nat}s_i$. Then $s\sim r$ as required. 
\end{proof}
}
\begin{proposition}\label{way-below-random}
    Suppose $r\in (\Omega\to D)$ is a random variable with a simple valuation $\alpha\ll T(r)$. Then there exists a simple random variable $s$ with $s\ll r$ and $T(s)=\alpha$.
\end{proposition}
\begin{proof}
Suppose $\beta=T(r)$ and $\alpha=\sum_{i\in I} p_i\delta(d_i)$, where $d_i$'s are assumed distinct for $i\in I$. By Proposition~\ref{simp-val-way}, for each $i\in I$, we have $p_i<\beta(\dua d_i)$, and thus, there exists a basic open subset $O_i\subseteq \Omega$ such that 
$O_i\ll_{\tau_A} r^{-1}(\dua d_i)$ and $\nu(O_i)=p_i$. Let $s=\sup_{i\in I}d_i\chi_{O_i}$. Then we have $T(s)=\alpha$, and, moreover, for each $i\in I$ we have $d_i\chi_{O_i}\ll r$ since $O_i\ll_{\tau_A} r^{-1}(\dua d_i)$. It follows that $s\ll r$.
\end{proof}

\begin{theorem}\label{random-valuation}
The map $T$ is a continuous function onto $PD$, mapping step functions to simple valuations. \end{theorem}
\begin{proof}
By Lemma~\ref{step-simple}, $T$ maps step functions to simple valuations. Monotonicity of $T$ is simple to check.  Suppose $(r_i)_{i\in I}$ is a directed set of random variables with $r=\sup _{i\geq I}r_i$ and $O\in \tau_D$ is a Scott open set. Then, by Lemma~\ref{map-frame-map},we have: 
\[r^{-1}(O)=\bigcup_{i\in I} r_i^{-1}(O).\]
Hence, since $\nu$ is a continuous valuation, we have:
\begin{multline}
(T(\sup_{i\in I} r_i))(O)=\nu((\sup_{i\in I} r_i)^{-1}(O))=\nu(\bigcup_{i\in I} r_i^{-1}(O)) \\
=\sup_{i\in I} \nu(r_i^{-1}(O))=\sup_{i\in I}(Tr_i)(O).
\end{multline}

To show that $T$ is onto, we first show that $T$ is onto the set of simple valuations with dyadic coefficients. Suppose $\alpha=\sum_{i=1}^nq_i\delta(d_i)$ is a simple valuation with $q_i$ a dyadic number and $d_i\in D$ for $1\leq i\leq n$ with $\sum_{i=1}^nq_i=1$. Since each $q_i$ is dyadic with $\sum_{i=1}^nq_i=1$, there exist disjoint basic open sets $(O_i)_{1\leq i\leq n}$ with $\sum_{i=1}^n\mu(O_i)=1$. Put $r=\sup_{1\leq i\leq n}d_i\chi_{O_i}$. Then $T(r)=\alpha$.

Suppose now  $\alpha\in PD$. Then there exists an increasing chain of simple valuations $(\alpha_i)_{i\geq 0}$ each with dyadic coefficients such that $\sup_{i\geq 0}\alpha_i=\alpha$. Using Lemma~\ref{eq-uniformity}, we can then inductively construct an increasing sequence of step functions $(r_i)_{i\geq 0}$ with $T(r_i)=\alpha_i$ for $i\geq 0$. By the continuity of $T$ we have: $T(\sup_{i\geq 0}r_i)=\sup_{i\geq 0}T(r_i)=\sup_{i\geq 0}\alpha_i=\alpha$.

\end{proof}
We note here that in~\cite[Corollary 4.10]{Mislove16}---using a sequence of results in topology, classical measure theory and domain theory---a proof is deduced that any continuous valuation on a bounded complete domain corresponds to a random variable on the domain with respect to the sample space $\Sigma^\nat$. In contrast, the proof of surjectivity of $T$ given in Theorem~\ref{random-valuation} is both direct and elementary. 
\subsection{Way-below preserving property}From the definition of $T$ we have: 
 $T(r_1)=T(r_2)$ iff $\nu(r_1^{-1}(O))=\nu(r_2^{-1}(O))$ for all Scott open sets $O\subseteq D$.
For the sample spaces denoted by $\Omega_0$, i.e., $\Sigma_0^\nat$ or $(0,1)$, we have the following additional and important property based on the lemma below.

\begin{lemma}\label{way-below-A-0}
   The collection $B_{(\Omega_0\to D)}$ of step functions that take the value $\bot$ in a non-empty open set is a basis for $(\Omega_0\to D)$.
\end{lemma}

\begin{proof}
  In fact, if $d\chi_O$, for $O\subseteq \Omega_0$ and $d\in D$, is any single step function then, since there are no compact-open sets in $\Omega_0$, there exists an increasing sequence of open sets $(O_i)_{i\in \nat}$ with $O_i\ll O$ satisfying $O=\bigcup_{i\in \nat}O_i$. Also, there is an increasing sequence $(d_i)_{i\in \nat}$ with $d_i\ll d$ satisfying $\sup_{i\in \nat}d_i=d$. Thus, $d_i\chi_{O_i}\ll d\chi_O$ for $i\in \nat$ with $\sup_{i\in \nat} d_i\chi_{O_i}=d\chi_O$. Since $O_i\ll O$ for $i\in \nat$, it follows that $d_i\chi_{O_i}\in B_{(\Omega_0\to D)}$.   
\end{proof}
It follows from Lemma~\ref{way-below-A-0} that, for simple valuations $r,s\in (\Omega_0\to D)$, the relation $r\ll s$ implies that $r$ takes values $\bot$ in a non-empty open set. Moreover, we have $T[B_{(\Omega_0\to D)}]=B_{PD}$ where $B_{PD}$ is given in Equation~(\ref{can-base}).
\begin{proposition}\label{random-simple-way-above}
For $b\in B_{(\Omega_0\to D)}$, we have $T[\dua b]=\dua (T(b))$.
\end{proposition}
\begin{proof}
  For $b\in B_{(\Omega_0\to D)}$, the set $\{y\in PD: T(b)\ll y\}$ is non-empty by Proposition~\ref{sp-way}. Let $b,b'\in B_{(\Omega_0\to D)}$ with $b\ll b'$. Then, by taking the interior of the crescents giving the values of $d$ and $b'$, we can write $b=\sup_{0\leq i\leq n}d_i\chi_{O_i}$ and $b'=\sup_{0\leq j\leq m}d_j'\chi_{O'_j}$ with $d_0=d'_0=\bot$, where $d_i$'s (respectively $d'_j$'s) are distinct, each $O_i$ for $0\leq i\leq n$, (respectively, each $O'_j$ for $0\leq j\leq m$) is a finite union of open intervals and $O_i$'s (respectively, $O_j'$'s) are pairwise disjoint. The value of $b$ (respectively, $b'$) in the finite set $\Omega_0\setminus\bigcup_{0\leq i\leq n}O_i$ (respectively, $\Omega_0\setminus\bigcup_{0\leq j\leq m} O'_j$) is given, as a result of Scott continuity, by the infimum of its values in the neighbouring open intervals.
Then, $b\ll b'$ implies that for each $i=0,\ldots,n$, we have:
\[O_i\ll \bigcup\{O_j:d_i\ll d'_j\}\]
We now define $t_{ij}$ as required in the splitting lemma to verify the way below relation $T(b)\ll T(b')$. We put:
\[  t_{ij} :=\left\{\begin{array}{ccc}
     \nu( O_i\cap O'_j)&\mbox{if } d_i\ll d'_j&i\neq 0\neq j\\
      \nu(O'_j\setminus \bigcup_{1\leq k\leq n}O_k)&&i=0\neq j\\
     \nu(O'_0)  &&i=0=j
\end{array}\right.\]
Note that $\nu(\bigcup_{0\leq i\leq n} O_i)=\nu(\bigcup_{0\leq j\leq m} O_j')=1$.  By a simple check, it follows that the splitting lemma conditions for $T(b)\ll T(b')$ in Proposition~\ref{sp-way} hold: 
\begin{itemize} 
\item For $i\neq 0$, $\nu(O_i)=\sum\{\nu(O_i\cap O'_j):d_i\ll d'_j,j\neq 0\}$\\ $=\sum\{t_{ij}:d_i\ll d'_j,j\neq 0\} $. 
\item For $i=0$, $\nu(O_0)=\nu(O')+\nu(\bigcup_{j\neq 0}O'_j\setminus\bigcup_{i\neq 0}O_i)$\\$=t_{00}+\sum\{t_{0j}:j\neq 0\}=\sum_{j}t_{0j}$.
\item For $j\neq 0$, we have,
$\nu(O_j')=\sum\{\nu(O'_j\cap O_i):d_i\ll d'_j\}=\sum\{t_{ij}:d_i\ll d'_j\}$.
\item Finally, $\nu(O_0')=t_{00}$.
\end{itemize}
Thus, $T(b)\ll T(b')$ and hence $T[\dua b]\subseteq \dua T(b)$. Now assume $b\in B$ and take any simple valuation $\alpha\in PD$ with $T(b)\ll \alpha$. By the comment after Lemma~\ref{eq-uniformity}, there exists $b'\in B$ such that $b\ll b'$ and $T(b')=\alpha$. Thus, $\dua T(b)\subseteq T(\dua b)$. 
\end{proof}
\begin{corollary}\label{way-below-preserve} The map $T:(\Omega_0\to D)\to PD$ preserves the way-below relation and is, thus, an open map.
\end{corollary}

\begin{corollary} \label{saturated-open} 
For any open set $U\subseteq (\Omega_0\to D)$ the set given by $\{r\in (\Omega_0\to D): \exists s\in U. \,r\sim s\}$ is open.
\end{corollary}
\begin{proof}
Since $\{r\in (\Omega_0\to D): \exists s\in U.\, r\sim s\} =T^{-1}(T[U])$ and $T[U]$ is open, the result follows from the  continuity of $T$.
\end{proof} 
For $\Omega = \Sigma^{\nat}$, Corollary~\ref{saturated-open} does not hold. A counterexample is given by taking two elements $d_1 \ll d_2 \in D$, the family of cylinder sets $C_i = [1^i0]$, where $1^i$ denotes a sequence of $1$ of lenght $i\in \nat$,  and the family of clopen sets $C'_j = \bigcup_{i=0}^j C_i$ (e.g.  $C'_2 = [0] \cup [10] \cup [110]$).  Consider the single step functions  $s_i = d_2 \chi_{C'_i}$ for $i\in \nat$. The random variables $s = \sup_{i\in \nat} s_i$ and $r_2 = d_2 \chi_{\Omega}$ are equivalent ($r_2 \sim s$). Consider now the open set $O = \dua(d_1 \chi_{\Omega})$ and its equivalence closure $O_c = \{r\in (\Omega\to D): \exists s\in O. \,  r\sim s\}$. We have $r_2 \in O$, therefore $s \in O_c$, while there is no $s_i$, for $i\in \nat$, contained in $O_c$.

\subsection{Computability of the probability map}

The relation $\sim$ on random variables is closed under supremum of increasing chains, reflecting Skorohod's Representation Theorem~\cite{billingsley2013convergence}:
\begin{proposition}\label{sup-closure}
 If $r_i,s_i\in \Omega \to D$ are two directed sets with $r_i\sim s_i$ for $i\in I$, then $\sup_{i\in I}r_i\sim \sup_{i\in I}s_i$.
\end{proposition}
\begin{proof}
  This follows from the continuity of $T$.   
\end{proof}

For $\Omega =[0,1]$ or $\Omega =\Sigma^\nat$ the map $T:(\Omega\to D)\to PD$ is not an open map. To see this take elements $d\ll d'$ with $d\neq \bot$, and let $r:=d\chi_{\Omega}$ and $s:=d'\chi_{\Omega}$. Then $r\ll s$ but $$T(r)=\delta(d)\ll\hspace{-1.5ex}\not\;\;\delta(d')=T(s)$$ by Proposition~\ref{sp-way}. However, we have the following result. 
 
\begin{theorem}\label{way-below-random-measure}
If $r\in (\Omega\to D)$ is a random variable, where $\Omega =\Sigma^\nat$ or $\Omega =[0,1]$, then there exists an increasing sequence of simple random variables $r_i$ with $r_i\ll r_{i+1}$ for $i\in \nat$, such that $T(r_i)\ll T(r)$ with $\sup_{i\in \nat} r_i=r$ a.e., with respect to $\nu$. 

\end{theorem}


\begin{proof}
Let $s_i$, for $i\in \nat$, be an increasing sequence of simple random variables with  $\sup_{i\in \nat} s_i=r$. Assume $s_i=\sup_{j\in I_i} d_{ij}\chi_{O_{ij}}$, for $i\in \nat$, where, for each $j\in I_i$, the elements $d_{ij}$ are distinct values of $s_i$ and the set $O_{ij}$ is a finite union of cylinders, when $\Omega =\Sigma^\nat$, or finite union of non-empty compact intervals, when $\Omega =[0,1]$. Take any single point $\omega_{ij}\in O_{ij}$ and let $O'_{ij}:=O_{ij}\setminus \{\omega_{ij}\}$ for $i\in \nat$ and $j\in I_i$ and put $r_i=\sup_{j\in I_i} d_{ij}\chi_{O'_{ij}}$. Then $r_i=s_i$ except at the finite set $\{\omega_{ij}:j\in I_i\}$ for each $i\in \nat$ and $r_i$ is an increasing sequence with, say, $r':=\sup_{i\in\nat }r_i$. Thus $r'$ is equal to $r$ everywhere except at the countable set of removed points $\bigcup\{\omega_{ij}:i\in \nat, j\in I_i\}$, i.e., $r'=r$ almost everywhere with respect to $\nu$. Since $O'_{ij}$ is not compact, for each $i\in \nat$ and $j\in I_i$, there exists an increasing sequence of clopen sets $O'_{ijn}$ with $O'_{ijn}\supset O'_{ij(n+1)}$ and $\nu(O'_{ijn})< \nu(O'_{ij(n+1)})$ for each $n\in \nat$ such that $O'_{ij}=\bigcup_{n\in \nat} O'_{ijn}$.   Take also an increasing sequence $d_{ijn}$ with $d_{ijn}\ll d_{ij(n+1)}$ for $n\in \nat$ such that $d_{ij}=\sup_{n\in \nat} d_{ijn}$. Then $r_{in}\ll r_{i(n+1)}\ll r_i$ for $i,n\in \nat$ and in addition, by Proposition~\ref{sp-way}, we have: \[T(r_{in})\ll T(r_{i(n+1)})\ll T(r_i)\sqsubseteq T(r),\] for $i,n\in \nat$. We claim that the family $\{r_{in}: (i,n)\in\nat^2\}$ is directed. Let $(i,n),(k,m)\in \nat^2$ with, say, $i\geq k$. Then, we have 
\[r_{km}\ll r_{k}\sqsubseteq r_i=\sup_{n\in \nat }r_{in},\]
and, thus, there exists $l\in\nat$ with $r_{km}\sqsubseteq r_{il}$. Then $r_{in}, r_{km}\sqsubseteq r_{ip}$  where $p=\max{\{n,l\}}$. Similarly, it follows that $T(r_{in})$ is a directed family with respect to $\ll$ for $(i,n)\in\nat^2$. Since the net $\{r_{in}: (i,n)\in\nat^2\}$ ordered with respect to $\ll$ is countable, we can find a cofinal subnet that is an increasing sequence of simple random variables $r_i$ with $r_i\ll r_{i+1}$ and $T(r_i)\ll T(r_{i+1})$ for $i\in \nat$ satisfying $\sup_{i\in \nat} r_i=r$ almost everywhere. 
\end{proof}


\begin{corollary}
  If $r\in (\Omega\to D)$ is a random variable, there exists an increasing sequence of simple random variables $r_i$ with $r_i\ll r_{i+1}$ for $i\in \nat$, such that $T(r_i)\ll T(r)$ with $\sup_{i\in \nat} r_i=r$ a.e., with respect to $\nu$.   
\end{corollary}
\begin{proof}
  It remains to show the result for $\Omega_0=\Sigma_0^\nat$  or $\Omega_0=(0,1)$. But this follows immediately by taking any increasing sequence of simple random variables $r_i$ with $r_i\ll r_{i+1}$ for $i\in \nat$, with $\sup_{i\in \nat} r_i=r$ and observing from Corollary~\ref{way-below-preserve} that $T(r_i)\ll T(r_{i+1})$ for $i\in \nat$.
\end{proof}
Suppose now $D\in \BC$ is an effectively given bounded complete domain with basis $B_0$ as described in Subsection~\ref{effect}. We use the basis $B_0$ to construct the the basis $B_{PD}$ of $PD$, given in Equation~(\ref{can-base}), and the basis $B_{(\Omega \to D)}$ of $(\Omega\to D)$, given in Equation~(\ref{simpl-random}). It follows by Proposition~\ref{simp-val-way} that the way-below relation on $B_{PD}$ is decidable. Similarly, by Lemma~\ref{way-below-func}, we can deduce that the way-below relation on $B_{(\Omega \to D)}$, as well as the binary lub operation on it, is decidable. Thus we obtain effective structures on $PD$ and $(Q\to D)$.   
It is now straightforward, using Theorem~\ref{random-valuation}, to deduce the following result.

\begin{corollary}
 The mapping $T:(\Omega\to D)\to PD$ is computable. Moreover, given a computable continuous valuation $\sigma\in PD$ and any simple valuation $\sigma_0\ll \sigma$, there is a computable random variable $r$ and a simple random variable $r_0\ll r$ such that $T(r)=\sigma$ and $T(r_0)=\sigma_0$.
\end{corollary}

\section{Monads for random variables}\label{monad-sec}

\subsection{PER-domains}

We next define a Cartesian closed category having a monad construction for random variables, this construction can then be used to give semantics to probabilistic functional programming languages. 

In the literature, the probabilistic monad is mostly defined as a space of probability measures, continuous valuations or the probabilistic powerdomain construction.  In all these cases the elements in the monad construction are functions assigning probability values to some selected sets of possible results for the computation. In our approach, the monad is defined as a space of random variables.


The equivalence relation $\sim$ between random variables induces a partial equivalence relation (PER), on the function spaces defined on domains of random variables. Therefore, we introduce the notion of PER-domains, i.e. domains with a partial equivalence relation on its elements. A further reason to introduce this new category of domains is that, as we will show, the monad diagrams commute only up to an equivalence relation on morphisms.

In the literature, there are several works where the notion of cpo with PER is introduced, \cite{AP90,Danielsson2006,sabelfeld2001per}. However, these works use rather different definitions and have different aims. 

\begin{definition}
A partial equivalence relation (PER), on a generic set, is a relation that is symmetric and transitive but not necessarily reflexive. A PER-domain $\langle D, \sim_D \rangle$ is a bounded complete domain, $D \in \BC$, with a partial equivalence relation $\sim_D$ on it which is also a logical relation~\cite{mitchell1989toward}, i.e. it satisfies the following two properties: 
\begin{itemize}
    \item $\bot \sim_D \bot$,
    \item for any pair of increasing chains $\langle d_i \rangle_{i \in N}$ and $\langle d'_i \rangle_{i \in N}$, if $\forall_i \ldot d_i \sim_D d'_i$ then $\sup_{i \in N} d_i \sim_D \sup_{i \in N} d'_i$.
\end{itemize}
Observe that by Proposition~\ref{sup-closure}, the equivalence relation for random variables on a domain satisfies the conditions of a logical relation, which is then extended in the above definition to PER-domains. We denote by $\PD$ the categories whose objects are PER-domains and whose 
morphisms are \emph{equivalence classes} of Scott-continuous functions between the underlying bounded complete domains, under the PER $\sim_{(D \to E)}$ defined by: 
\[ f_1 \sim_{(D \to E)} f_2 \mbox{ iff } d_1 \sim_D d_2\, \mbox{implies}\ f_1 (d_1) \sim_E f_2(d_2).
\]
Composition of morphisims is defined by $[f] \circ [g] = [f \circ g]$
\end{definition}

Since composition preserves the PER relation on morphisms, the above definition is well-posed.  
Notice that in writing $[f]$, we implicitly assume that $f$ defines a non-empty equivalence class, so in particular $f$ should be equivalent to itself, that is $f$ preserves the partial equivalence relation.

It is immediate that any bounded complete domain can be considered a PER-domain with the partial equivalence relation defined as equality. It follows that there is an obvious faithful functor from the category $\BC$ to the category of PER-domains. 


The standard domain constructions are extended on PER-domains using the standard definition for logical relations:
\begin{definition}
Given two PER-domains $\langle D, \sim_D \rangle$  and $\langle E, \sim_E \rangle$, the product PER-domain consists of the domain $D \times E$ with a PER defined by: $(d_1, e_1) \sim_{D \times E} (d_2, e_2)$ iff $d_1 \sim_D d_2$ and $e_1 \sim_E e_2$.
The function space PER-domain consists of the domain $(D \to E)$ with a PER defined by: $f_1 \sim_{(D \to E)} f_2$ iff for every $d_1 \sim_D d_2$, we have: $f_1 (d_1) \sim_E f_2(d_2)$.
\end{definition}

\begin{proposition}
$\PD$ is a Cartesian closed category.
\end{proposition}
\begin{proof}
    Every projection, for example $\pi_1 : (D \times E) \to D$, preserves the equivalence relation so it defines an equivalence class $[\pi_1] : \ \langle D \times E, \sim_{D \times E} \rangle \to \langle D, \sim_D \rangle$.  It is also simple to check that for any pair of morphisms \[[f] : \langle C, \sim_C \rangle \to \langle D, \sim_D \rangle,\qquad[g] : \langle C, \sim_C \rangle \to \langle E, \sim_E \rangle,\] the function $\langle f, g \rangle : C \to (D \times E)$ preserves the PER, and so $[\langle f, g \rangle]$ is the unique morphism in $\PD$ making the diagram for Cartesian product commute. 

    Given $f_1, f_2$ in $(D \times E) \to F$, we have $f_1 \sim f_2$ iff the corresponding curryied functions
    $f'_1, f'_2$ in $D \to (E \to F)$ are equivalent, i.e., $f'_1 \sim f'_2$, and it follows that there exists a bijection between the equivalence classes in  $(D \times E) \to F$ and in $D \to (E \to F)$ inducing a natural transformation.
\end{proof}


\subsection{Monad construction and R-topology}
We now aim to define a probabilistic monad for the category $\PD$. To this effect, we need to define a topology on any PER domain $\langle D,\sim_D\rangle$ which is weaker than the Scott topology on $D$.
\begin{definition}
We say a Scott open subset $O\subseteq D$ in a PER-domain $\langle D,\sim_D\rangle$ is {\em R-open} if it is closed under $\sim_D$, i.e., $d_1 \in O$ and $d_1\sim d_2$ implies $d_2\in O$. \remove{The R-topology $\Upsilon_D$ on a PER-domain $D$ is the sub-topology of the Scott topology such that $(d_1\sim_D d_2)\iff \forall O\in \Upsilon_D. (d_1\in O\iff d_2\in O)$.}
\end{definition} 
It is easy to check that the collection of all R-open sets of a PER-domain $\langle D,\sim_D\rangle$ is a topology, i.e., R-open sets are closed under taking arbitrary union and finite intersections. We call this topology the {\em R-topology} on $\langle D,\sim_D\rangle$.

The {\em random variable functor} $R$ is defined as follows:

\begin{definition}\label{def-monad}
The functor $R:\PD \to \PD$ on the object $\langle D, \sim_D \rangle \in \PD$ is defined by \[R \langle D, \sim_D \rangle= \langle (\Omega\to D), \sim_{R \, D} \rangle,\]with $r_1 \sim_{R \, D} r_2$ if
\begin{itemize}
    \item for any $\omega \in \Omega$, $r_1(\omega) \sim_D r_1(\omega)$ and $ r_2(\omega) \sim_D r_2(\omega)$,
    \item for any R-open set $O\subseteq D$, we have: $\nu(r_1^{-1}(O)) = \nu(r_2^{-1}(O))$.
\end{itemize}
  On morphism $[f]: \langle D, \sim_D  \rangle \to \langle E, \sim_E \rangle$,  the functor is defined as: \[R[f]: \langle (\Omega\to D), \sim_{RD}  \rangle \to \langle (\Omega\to E), \sim_{RE} \rangle,\] given by $R[f] = [ \lambda r \, . f \circ r]$. 
\end{definition}

\begin{proposition}
    The functor $R : \PD \to \PD$ is well-defined.
\end{proposition}
\begin{proof}
We need to show that if $f_1 \sim_{D \to E} f_2$
then $\lambda r . f_1 \circ r \sim_{RD \to RE} \lambda r . f_2 \circ r$, which in turn amounts to showing that for any  
$r_1 \sim_{R D} r_2$, we have: $f_1 \circ r_1 \sim_{R E} f_2 \circ r_2$.   Equivalently,  for any open set $O$ closed under $\sim_E$, we have:
$\nu ((f_1 \circ r_1)^{-1} (O)) = \nu ((f_2 \circ r_2)^{-1} (O))$, or more explicitly, 
$\nu (r_1^{-1} (f_1^{-1} (O))) = \nu (r_2^{-1} (f_2^{-1} (O)))$.
Now, since $O$ is closed under $\sim_E$, we have that $f_2^{-1}(O)$ is closed under $\sim_D$ and for any 
$x \sim_D x$, $x \in f_1^{-1}(O)$ iff $x \in f_2^{-1}(O)$.
Since the image of $r_1$ contains only elements equivalent to themselves, we can write $\nu (r_1^{-1} (f_1^{-1} (O))) = \nu (r_1^{-1} (f_2^{-1} (O)))$ which, since $r_1 \sim r_2$ and  $f_2^{-1}(O)$ is closed under $\sim_D$, in turn is equal to $\nu (r_2^{-1} (f_2^{-1} (O)))$. 
\end{proof}

It follows that any function $f : D \to E$ can be lifted to the PER-domain of random variables \[R[f]: \langle (\Omega\to D), \sim_{RD}  \rangle \to \langle (\Omega\to E), \sim_{RE} \rangle,\] by first applying the faithful immersion of domains in PER-domains and then applying the functor $R$. In particular, we will have the following result by assuming $D=E=\realDom$ and considering arithmetic and more generally any operations on random variables on $\realDom$
\begin{corollary}\label{multi-distribution}
The pointwise application of the basic arithmetic operations and of any continuous function on random variables 
defines self-related maps. 
\end{corollary}
Corollary~\ref{multi-distribution} is easily extended to higher order types, built from $\realDom$ using the function space, product and the monad functors. 








Next, we show that $R$ induces a monad.  
The monad construction uses as parameters the completion, equivalent in our case with the closure  $\overline{\Omega}$ of $\Omega$ and a function $h: \overline{\Omega}\to \overline{\Omega}\times \overline{\Omega}$, which is
continuous, surjective and injective on a subset of $\Omega$ with full measure, i.e. measure $1$, with the additional condition that the map $h$ is measure preserving by taking the product measure $\nu \times \nu$ on $\overline{\Omega}\times\overline{\Omega} $.

We then define $\Omega' = \Omega \cap h^{-1}(\Omega \times \Omega) $ and $h_1, h_2 : \overline{\Omega} \to \overline{\Omega}$ as $h_1 = \pi_1 \circ h$ and $h_2 = \pi_2 \circ h$.  Note that ${\nu}(\Omega') = 1$ and that $h_1$ and $h_2$ are also measure preserving since $\pi_1$ and $\pi_2$ are measure preserving and the composition of measure preserving maps is measure preserving. Since $h: \overline{\Omega}\to \overline{\Omega}\times \overline{\Omega}$ is measure preserving and injective almost everywhere, on $\overline{\Omega}\times\overline{\Omega} $ the push-forward measure $\nu\circ h^{-1}$ and the product measure $\nu\times \nu$ coincide: $\nu\circ h^{-1}= \nu\times \nu$.


There are infinitely many monads one can obtain by choosing such  $\overline{\Omega}$ and $ h$. We present four canonical cases in the following.
\begin{definition}
    \begin{itemize} 
\item[(i)] Take $\Omega = \Sigma^\nat$ or $\Omega =\Sigma^\nat_0$ with
$\overline{\Omega}= \Sigma^\nat$, and $h$ defined by $h(\omega) = (\omega^{\operatorname{e}}, \omega^{\operatorname{o}})$, where $\omega^{\operatorname{e}}$ (respectively, $\omega^{\operatorname{o}})$ is the sequence of values in even (respectively, odd) positions in the sequence $\omega$, i.e., $(\omega^{\operatorname{e}})_i=\omega_{2i}$ and $(\omega^{\operatorname{o}})_i=\omega_{2i+1}$ for $i\in \nat$. Then the inverse $k:=h^{-1}:\Omega^2\to \Omega$ is given by: $k(\omega,\omega')=\omega''$, where $\omega''_{2i}=\omega_i$ and $\omega''_{2i+1}=\omega'_i$ for $i\in \nat$.
\item[(ii)] Take $\Omega =[0,1]$ or $\Omega =(0,1)$ with $\overline{\Omega}= [0,1]$, and the map $h$ given by the Hilbert curve \cite{sagan2012space} as described in Subsection~\ref{hilbert-sec}.
\end{itemize}
\end{definition}

The closure $\overline{\Omega}$ of the sample space $\Omega$ is necessary to accommodate as possible sample spaces $\Sigma^\nat_0$ and $(0,1)$ for which there is no simple continuous, surjective and measure preserving function $h : \Omega \to \Omega^2$.  

\subsection{Monadic properties}
 In this section, we show that $R$ defines a monad. 
 
 Let $\eta_D : D\to RD$ be given by $\eta_D(d)(\omega) = d$ and 
 \[\mu_D:(\Omega \to \Omega\to D)\to \Omega \to D,\] be given by $\mu_D (r)(\omega)= r^\star (h_1 (\omega)) (h_2 (\omega))$.  Where $r^\star$ is the envelope of the function $r$ as defined in Proposition~\ref{envelope}.
For an input $\omega \in \Omega'$, instead of the envelope $r^\star$, we can use $r$ itself and the simpler equation $\mu_D (r)(\omega)= r (h_1 (\omega)) (h_2 (\omega))$  holds.
 
Formally we define 
$\eta_{\langle D, \sim_D\rangle} : \langle D, \sim_D\rangle
\to R\langle D, \sim_D\rangle$ as 
$\eta_{\langle D, \sim_D\rangle} = [\eta_D]$ and similarly for 
$\mu_{\langle D, \sim_D\rangle}$
To verify that $\eta_D$ and $\mu_D$ induce morphisms on PER-domains, we need to show they preserve the equivalence relation. For $\eta_D$ this is immediate, while for $\mu_D$, we require some preliminary results. 

\begin{definition}
  Given an R-open set $O\subseteq D$ of $\langle D,\sim_D\rangle$ and a real number $0\leq q\leq 1$, the subset $[q\to O]$ of $\langle RD,\sim_{RD}\rangle$
  is defined by
  \[[q\to O]:= 
  \{ r: (\Omega \to D) : 
  \nu(r^{-1}(O))>q\}. \]
\end{definition}
Note  from the definition that $[0\to D]=RD$ and, for any R-open set $O$, we have: $[1\to O]=\emptyset$ since $\nu(\Omega)=1$.
\begin{lemma}\label{is-scott-open}
The set $[q\to O]\subseteq (\Omega\to D)$ is R-open for any R-open set $O\subseteq D$, for all $q\in [0,1]$. 
\end{lemma}
\begin{proof}
    We first check that $[q\to O]$ is Scott open. Clearly, $[q\to O]$ is upper closed.  Suppose now that $r_n\in (\Omega\to D)$ is an increasing sequence with $r=\sup_{n\in\nat}r_n\in [q\to O]$. Then, by Lemma~\ref{map-frame-map}, we have: $r^{-1}(O)=\bigcup_{n\in \nat}r_n^{-1}(O)$.
Since $r_n^{-1}(O)$, for $n\in \nat$, is an increasing sequence of opens with $\bigcup_{n\in \nat}r_n^{-1}(O)>q$, there exists $n\in \nat$ with $\nu(r_n^{-1}(O))>q$, i.e., $r_n\in [q\to O]$.  Thus $[q\to O]$ is Scott open. If $r_1\in [q\to O]$ and $r_1\sim_{RD} r_2$ then, by definition, $\nu(r_2^{-1}(O))= \nu(r_1^{-1}(O))>q$, i.e., $r_2\in [q\to O]$ and the result follows. 
\end{proof}
\remove{The following two examples show that, for $\Omega =\Sigma^\nat$, the two conditions in Definition~\ref{def-monad} of $R$ are independent.  Recall that $[x_1x_2\ldots x_n]\subseteq 2^\nat$ denotes the clopen set of infinite sequences with initial segment $x_1x_2\ldots x_n$.

\begin{itemize} 
    \item[(i)] Let $F,G:\Sigma^\nat\to (\Sigma^\nat\to \realDom)$ be given by 
    \[F=\sup\{([0,1]\chi_{[0]})\chi_{[0]},([-1,0]\chi_{[0]})\chi_{[1]}\}\]
    \[G=\sup\{([-1,0]\chi_{[0]})\chi_{[0]},([0,1]\chi_{[0]})\chi_{[1]}\}\]
    Then, we have $ \nu\circ F^{-1}=\nu\circ G^{-1}$; however, $F(\omega)\sim G(\omega)$ does not hold for any $\omega\in 2^\nat$.
    \item [(ii)] Let $F,G:\Sigma^\nat\to (\Sigma^\nat\to \realDom)$ be given by \[F=\sup \{\sup (A_1)\chi_{[00]},\sup (A_2)\chi_{[1]\cup [01]}\}\]  \[G=\sup \{\sup (B_1)\chi_{[0]},\sup (B_2)\chi_{[1]}\}\] where 
    \[A_1=\{[0,1]\chi_{[0]},[-1,0]\chi_{[1]}\} \]
    \[A_2= \{[-1,0]\chi_{[0]},[0,1]\chi_{[1]}\}\]
     \[B_1=\{[-1,0]\chi_{[0]},[0,1]\chi_{[1]}\}\]
     \[B_2=\{[0,1]\chi_{[0]},[-1,0]\chi_{[1]}\}\]
     Then $F(\omega)=G(\omega)$ for $\omega\in 2^\nat$ but $\nu\circ F^{-1}\neq \nu\circ G^{-1}$.
\end{itemize}}
\begin{lemma}\label{lem-basic-eq}
If $D$ is a PER domain and $r:\Omega \to (\Omega\to D)$ a random variable and $O\subseteq D$ an R-open set, then we have:
\begin{equation}\label{basic-eq}
\int_{\Omega}\nu((r(\omega))^{-1}(O))\,d\omega =\int_0^1\nu(r^{-1}([q\to O]))\,dq\end{equation}
\end{lemma}
\begin{proof}
Since $r$ is Scott continuous, the integrands \[\omega\mapsto \nu((r(\omega))^{-1}(O)):\Omega \to \realLine,\]\[q\mapsto \nu(r^{-1}([q\to O]):[0,1]\to \realLine,\] are both bounded upper-continuous functions and thus Lebesgue integrable. By Lebesgue's monotone convergence theorem~\cite{walter1987real}, it is sufficient to show the equality for a step function $r$. Assume
\[r=\sup_{i\in I}\left (\left(\sup_{j\in I_i}d_{ij}\chi_{V_{ij}}\right)\chi_{U_i},  \right)\]
where $I$ and $I_i$ for $i\in I$ are finite indexing sets, $U_i\subseteq \Omega$ for $i\in I$ are disjoint crescents, for fixed $i\in I$, the crescents $V_{ij}$ for $j\in I_i$ are disjoint and $d_{ij}\in D$ for $i\in I$ and $j\in I_i$. We now compute the LFH and the RHS of Equation~(\ref{basic-eq}).

LHS: We have $r(\omega)=\sup\{d_{ij}\chi_{V_{ij}}\}$ for $\omega \in U_i$. Thus, \\
$(r(\omega))^{-1}(O)=\bigcup_{d_{ij}\in O}V_{ij}$ for $\omega\in U_i$. Therefore,
\[\int_A\nu((r(\omega))^{-1}(O)\,d\omega=\sum_{i\in I}\nu(U_i)\left(\sum\nu(V_{ij}):j\in I_i,d_{ij}\in O\right)\]
RHS: We first note the following equality: $r^{-1}([q\to O]) =$ \\ $\bigcup_{i\in I}\{\omega\in U_i: \sup_{j\in I_i} d_{ij}\chi_{V_{ij}}\in [q\to O]\}$. In other words, 
\[r^{-1}([q\to O])=\bigcup_{i\in I}\left\{\omega\in U_i: \sum_{j\in I_i} \nu\left(\bigcup_{d_{ij}\in O}V_{ij}\right)>q\right\}.\]
Hence, putting \[q_i:=\sum_{j\in I_i} \nu\left(\bigcup_{d_{ij}\in O}V_{ij}\right)=\sum_{j\in I_i,d_{ij}\in O}\nu(V_{ij}),\] for each $i\in I$, we have:
\[\int_0^1\nu(r^{-1}([q\to O]))\,dq \] 
\[ =\int_0^1\sum_{i\in I} \nu\left(\left\{\omega\in U_i \mid \sum_{j\in I_i} \nu\left(\bigcup_{d_{ij}\in O}V_{ij}\right)>q\right\}\right)\,dq \]
\[=\sum_{i\in I} \nu(U_i)\left(\int_0^{q_i} \,dq\right)=\sum_{i\in I} \nu(U_i)q_i \]
\[=\sum_{i\in I} \nu(U_i)\left (\sum \nu(V_{ij}):j\in I_i, d_{ij}\in O\right).\]
Thus LHS and RHS coincide and the result follows. 
\end{proof}
\begin{lemma}\label{currying}
   If $r_1\sim_{\Omega \to (\Omega\to D)}r_2$, then for all R-open sets $O\subseteq D$ of the PER-domain $\langle D,\sim_D\rangle$, we have: \[(\nu\times \nu)(r'^{-1}_1(O))=(\nu\times\nu)(r'^{-1}_2(O)),\] where $r'_i(\omega_1,\omega_2)=(r_i(\omega_1))(\omega_2)$ for $i=1,2$.
\end{lemma}
\begin{proof}

If $r:\Omega \to (\Omega\to D)$ is Scott continuous then the map $\nu((r(-))^{-1}(O)): \Omega \to \realLine$ with $\omega\mapsto\nu((r(\omega))^{-1}(O))$ is bounded and upper semi-continuous by the Scott conintuity of $r$, and is thus Lebesgue integrable. Similarly, the map \[(\nu\times \nu)(r'(-,-))^{-1}(O): \Omega \times \Omega \to \realLine,\] \[(\omega_1,\omega_2)\mapsto (\nu\times\nu)((r'(\omega_1,\omega_2)^{-1}(O)),\] is Lebesgue integrable, where $r'(\omega_1,\omega_2)=(r(\omega_1))(\omega_2)$.
We have by first invoking Fubini's theorem, then using Lemma~\ref{lem-basic-eq} and finally the relation $r_1\sim r_2$: 
\[(\nu\times \nu)(r_1'^{-1}(O))=\int_{\Omega\times \Omega}(\nu\times \nu)((r_1'(\omega_1,\omega_2))^{-1}(O))\,d\omega_1d\omega_2\]
\[=  \int_{\Omega}\nu\left(\int_{\Omega}\left(\nu((r_1'(\omega_1,\omega_2))^{-1}(O)\right)\,d\omega_2\right)d\omega_1 \] 
\[=\int_{\Omega}\nu((r_1(\omega_1))^{-1}(O))\,d\omega_1=\int_0^1r_1^{-1}([q\to O])\,dq,\]
\[=\int_0^1r_2^{-1}([q\to O])\,dq=(\nu\times \nu)(r_2'^{-1}(O))\]
\end{proof}
The opposite implication in Lemma~\ref{currying} is not true, i.e. there exists a pair of random variables $r_1, r_2 \in (\Omega \to (\Omega\to D))$ such that $r_1 \not \sim_{\Omega \to (\Omega\to D)}r_2$ although for any open set $O\subseteq D$ of the PER-domain $\langle D,\sim_D\rangle$ the equality $(\nu\times \nu)(r'^{-1}_1(O))=(\nu\times\nu)(r'^{-1}_2(O))$ holds.  A simple counterexample can be obtained by taking $\Omega =[0,1]$ and $D=\interval[0,1]$ and defining $r_1(x)(y) = [x,x]$ and $r_2(x)(y) = [y,y]$. Given the open subset of $([0,1] \to \interval[0,1])$ defined by $O = [ 1/3 \to \dua [0,1/2]]$, we have that $\nu (r_1^{-1}(O)) = 1/2$ and $\nu (r_2^{-1}(O)) = 1$. While for any open set $O'\subseteq \interval[0,1]$, clearly: $(x,y) \in r'^{-1}_1 (O')$ iff $(y,x) \in r'^{-1}_2(O'))$, and therefore $\nu (r'^{-1}_1(O') = \nu( r'^{-1}_2(O')))$.

\begin{lemma}
    $\mu_D(r_1)\sim_{\Omega\to D}\mu_D(r_2)$ if $r_1\sim_{\Omega \to (\Omega\to D)}r_2$.
\end{lemma}
\begin{proof}
 
Let $r'_i$ for $i=1,2$ be as defined in Lemma~\ref{currying}.
For $i=1,2$ and for any R-open set $O$ we can write:
\[\begin{array}{ll} \nu ((\mu_D(r_i))^{-1}(O))&\\
=\nu(\{\omega \in \Omega:(r^\star_i(h_1(\omega)))(h_2(\omega))\in O\})&\\
=\nu(\{\omega \in \Omega':(r_i(h_1(\omega)))(h_2(\omega))\in O\})& \mbox{since } \nu(\Omega')=1 \\
=\nu(\{\omega \in \Omega':r'_i(h_1(\omega), h_2(\omega))\in O\})&\\
=\nu(h^{-1}(r_i'^{-1}(O)).
\end{array}\] 
Since,  by Lemma~\ref{currying},  $\nu(r_1'^{-1}(O)) = \nu(r_2'^{-1}(O))$ and $h^{-1}$ preserves measure,
we have that for any R-open set $O$,
$\nu ((\mu_D(r_1))^{-1}(O)) = \nu ((\mu_D(r_2)^{-1}(O))$ which conclude the proof.
\end{proof}
 
 \begin{proposition}
$(R,\eta,\mu)$ is a monad on $\PD$. 
 \end{proposition}
 \begin{proof}
  That $\eta$ gives a natural transformation is trivial to check. To check that $\mu$ is a natural transformation on $\PD$, we need to show that for any $[r]: \langle \Omega \to \Omega \to D, \sim \rangle$ and $[f]: \langle D \to E, \sim \rangle$, we have $R f\circ \mu_D(r) \sim_{\Omega\to E} \mu_E \circ RRf(r)$. 
  \[ \begin{tikzcd}[row sep=huge, column sep=huge]
\langle \Omega \to \Omega \to D, \sim \rangle \arrow{r}{\mu_{\langle D, \sim \rangle}} 
\arrow[swap]{d}{RR[f]} & 
\langle \Omega \to D, \sim \rangle \arrow{d}{R[f]} \\%
\langle \Omega \to \Omega \to E, \sim \rangle \arrow{r}{\mu_{\langle E, \sim \rangle}}& 
\langle \Omega \to E, \sim \rangle
\end{tikzcd}
\]
 On the one hand, for any $\omega \in \Omega'$ we can write: 
\[\begin{array}{l}(R  f\circ \mu_D)(r)(\omega) = (R f (\mu_D (r))(\omega)
 \\
= f ((\mu_D (r))(\omega))= f (r (h_1 (\omega))(h_2 (\omega)))\end{array}.\] 

and on the other hand for any $\omega \in \Omega'$:
\[\begin{array}{l}(\mu_E \circ  R Rf)(r)(\omega) = \mu_E ((R Rf)(r))(\omega) \\
= (R R f) (r) (h_1(\omega))(h_2 (\omega)) = (R f) (r (h_1(\omega)))(h_2 (\omega)) \\
=  f (r (h_1 (\omega))(h_2 (\omega))).\end{array}\]
Since  $(R  f\circ \mu_D)(r)$ and $(\mu_E \circ  R Rf)(r)$ are equal a.e., they are equivalent, and the diagram commutes.
 
Next we check the cummutativity of the following diagram:
\[ \begin{tikzcd}[row sep=huge, column sep=large]
R^3 \langle D, \sim \rangle \arrow{r}{\mu_{R \langle D, \sim \rangle}} \arrow[swap]{d}{R\mu_{\langle D, \sim \rangle}} & 
R^2 \langle D, \sim \rangle \arrow{d}{\mu_{\langle D, \sim \rangle}} \\%
R^2 \langle D, \sim \rangle\arrow{r}{\mu_{\langle D, \sim \rangle}}& 
R \langle D, \sim \rangle
\end{tikzcd}
\]

Let $\Omega'' := \Omega \cap h^{-1}(\Omega' \times \Omega')$, for $r: R^3 D$, we have: 
  \[\mu_{R D}(r)(\omega_1)(\omega_2) 
  = (r (\omega_1))^\star (h_1 \omega_2) (h_2 \omega_2),\] 
  and thus, for $\omega \in \Omega'' $,
  \[ \mu_D(\mu_{R D}(r))(\omega) = r(h_1 \omega)(h_1 (h_2 \omega))(h_2 (h_2 \omega)).\]

On the other hand, 
\[\begin{array}{l}R\mu_D(r)(\omega_1)(\omega_2)\\
=  (R (\lambda s \omega .\, s (h_1 \omega)(h_2 \omega)))(r^\star)(\omega_1)(\omega_2) \\ [1ex]
= (\lambda s \omega .\, s (h_1 \omega)(h_2 \omega))(r^\star)(\omega_1)(\omega_2) \\ [1ex]
= r^\star (h_1 \omega_1)(h_2 \omega_1)(\omega_2),\end{array}\]
 and thus, for $\omega \in \Omega'' $,
 \[\mu_D(R\mu_D (r))(\omega)
 = r (h_1 (h_1 \omega))(h_2 (h_1 \omega))(h_2 \omega). \]

Let the functions $j, k : \Omega \to \Omega^3$, be defined by
\[j(\omega) := ((h_1 \omega),(h_1 (h_2 \omega)),(h_2 (h_2 \omega)))\]
\[k(\omega) := ((h_1 (h_1 \omega)),(h_2 (h_1 \omega)),(h_2 \omega)),\]
 which characterise the behaviour of $\mu_D(\mu_{R D}(r))$ and $\mu_D(R\mu_D (r))$. They
can also be defined by $j = [id,h] \circ h$ and $k = [h,id] \circ h$.  Since these are compositions of measure preserving, and almost everywhere bijective functions, $j, k$ are also measure preserving and almost everywhere bijective functions. 
It follows that there exists a partial function $k^{-1} : \Omega^3 \to \Omega$ such that almost everywhere \[(\omega_1, \omega_2, \omega_3) = (k \circ k^{-1})(\omega_1, \omega_2, \omega_3),\] i.e. $k^{-1}$ is  the inverse of $k$. 
It follows that, almost everywhere,  $\mu_D(R\mu_D (r)) = \mu_D(\mu_{R D}(r))  \circ k^{-1} \circ j$, from which it is easy to derive
$\mu_D(\mu_{R D}(r)) \sim_{RD} \mu_D(R\mu_D (r))$.   



Following a similar schema, next, we show the right and the left triangles commute in the diagram below. 
\begin{center}
\begin{tikzcd}[row sep=huge, column sep=huge]
R{\langle D, \sim \rangle} \arrow[rd, "\operatorname{id}_{R {\langle D, \sim \rangle}}" '] \arrow[r, "\eta_{R{\langle D, \sim \rangle}}"] & R^2{\langle D, \sim \rangle}\arrow[d,"\mu_{\langle D, \sim \rangle}"]& \arrow[ld, "\operatorname{id}_{R{\langle D, \sim \rangle}}"] \arrow[l,"R\eta_{\langle D, \sim \rangle}" ']R{\langle D, \sim \rangle}\\
& R{\langle D, \sim \rangle}&
\end{tikzcd}
\end{center}
Let $r:\Omega \to D$.  Then, for the right triangle we have:  \[(R\eta_D(r))(\omega_1)(\omega_2) = \eta_D(r) (\omega_1) (\omega_2)= r (\omega_2).\] Thus, $\mu_D(R\eta_D(r))= r \circ h_2$.  Since $h_2$ is measure-preserving, $r \sim_{R D} r \circ h_2$. 
Similar arguments apply for the left triangle, in fact: 
\\ 
$\eta_{\Omega\to D}(r)(\omega_1)(\omega_2)= r (\omega_1)$, and thus $\mu_D(\eta_{\Omega\to D}(r))= r \circ h_1$.
 \end{proof}

Alternatively, we have a Kleisli triple $(R,\eta,(-)^\dagger)$ as follows. Given $[f]:{\langle D, \sim \rangle}\to R{\langle E, \sim \rangle}$, we define $[f]^\dagger:R{\langle D, \sim \rangle}\to R{\langle E, \sim \rangle}$ as $[f]^\dagger = [f^\dagger]$ where $f^\dagger(r)(\omega) = f(r (h_1 (\omega))(h_2 (\omega)))$. Since $f^\dagger=\mu_{D}\circ Rf$, the  construction is correct.

\remove{
\begin{corollary}
If $f_1 \sim_{D \to RE} f_2$ and $r_1 \sim_{R D} r_2$ we have that $f_1^\dagger(r_1) \sim_{RE} f_2^\dagger(r_2)$.  
\end{corollary}
\begin{proof}
We simply recall that $f^\dagger=\mu_{D}\circ Rf$ and composition preserves $\sim$
\end{proof}

 **********************
 
\begin{proposition}
  $(R,\eta,(-)^\dagger)$ is a Kleisli triple.
\end{proposition}

\begin{proof}

(i)  If $f:D\to RE$, then for $x\in D$, we have: \\
$(f^\dagger\circ \eta_D)(d)(u)= 
  f (\eta_D(d)(h_1 (u)))(h_2 (u))= f(d)(h_2 (u))$, and $f(d) \sim_{R E} f(d) \circ h_2$.

\footnote{something more needs to be stated}
  
(ii) $\eta_D^\dagger(d)(u) = d(h_2 (u))$

(iii) Suppose $f:B\to RC$ and $g:C\to RD$. If $r\in RB$, then
\begin{align*}
    & g^\dagger(f^\dagger(r))(u)= g((f^\dagger(r))(h_1 (u))(h_2 (u)) \\
    & = g(f(r (h_1 \circ h_1(u)))(h_2 \circ h_1(u)))(h_2 (u)) 
\end{align*}
while 
\begin{align*}
 & (g^\dagger\circ f)^\dagger(r)(u) = (g^\dagger\circ f)(r (h_1 (u)))(h_2 (u)) \\ 
 & = g (f(r (h_1 (u)))(h_1 \circ h_2 (u)))(h_2 \circ h_2 (u))
\end{align*}
\end{proof}
}

\subsection{Commutativity of the monads}

We next show that $R$ is a strong commutative monad, for which we will explain and use the notions, properties and results presented in~\cite{moggi1991notions} as follows. In a category with finite products and enough points the monad $R$ is a strong monad if there are morphisms \[t_{D,E}:D\times RE\to R(D\times E),\] where $t$ is called tensorial strength, such that \begin{equation}\label{tensorial} t_{D,E}\circ \langle x,y\rangle=R(\langle x\circ !_E,\id_E\rangle)\circ y,\end{equation} 
where $!_E:E\to 1$ is the unique morphism from $E$ to the final object $1$. The dual tensorial strength $t'$ is given by a family of morphisms \[t'_{D,E}:RD\times E\to R(D\times E)\] whose action is obtained by swapping the two input arguments, applying $t_{D,E}$ and then swapping the arguments of the output. The monad $R$ is called commutative if the two morphisms $t'^\dagger_{D,E}\circ t_{RD,E}$ and $t^\dagger_{D,E}\circ t'_{D,RE}$ coincide. 

\begin{proposition}
$R$ is a strong commutative monad on $\PD$. 
\end{proposition}
\begin{proof}
Being Cartesian, the category $\PD$ has finite products.  Moreover $\PD$ has enough points: suppose the two morphisms \[[f], [g]: \langle D, \sim_D \rangle \to \langle E, \sim_E \rangle\] are equal on all points. Since points in $[x] :1\to \langle D, \sim_D \rangle$ are in one-to-one correspondence with equivalence classes $[d]$ in $\langle D, \sim_D \rangle$, it follows that $f(d) \sim_E g(d)$ for any $d \in D$ with $d \sim_D d$. More generally, for any pair $d_1 \sim d_2$, since $f \sim_{D \to E} f$, one has 
$f(d_1) \sim_E f(d_2)$, and similarly for $g$; by transitivity of $\sim_E$, it follows that $f(d_1) \sim_E g(d_2)$, and therefore  $[f] = [g]$.

For $\langle D, \sim_D \rangle, \langle E, \sim_E \rangle\in \PD $, we define the morphism \[t_{D,E}:D\times RE\to R(D\times E),\]by $t_{D,E}(d,s)=\langle \eta_D(d),s\rangle$. A simple calculation shows that $t_{D,E} \sim t_{D,E}$ and that $[t_{D,E}]$ satisfies the required condition in Equation~(\ref{tensorial}).  Hence, $R$ is a strong monad.

We also have $t'_{D,E}:RD\times E\to R(D\times E)$ given by $t'_{D,E}(r,e)=\langle r,\eta_E(e)\rangle$ and:
\[t'^\dagger_{D,E}\circ t_{RD,E}(r,s)=t'^\dagger_{D,E}\langle \eta_D(r),s\rangle=\langle \eta_D(r),\eta_E(s)\rangle.\]
By symmetry:
\[t^\dagger_{D,E}\circ t'_{D,RE}(r,s)=t^\dagger_{D,E}\langle r,\eta_E(s)\rangle=\langle \eta_D(r),\eta_E(s)\rangle.\]
Hence, $R$ is a strong commutative the monad. 
\end{proof}

\section{Random variables on $\realLine^n$}\label{rv-r} We now consider random variables on finite dimensional Euclidean spaces. We start by observing that the uniform distribution, i.e., Lebesgue measure, on $[0,1]$ is clearly represented by the identity map $\operatorname{Id}:[0,1]\to [0,1]$. In general, we construct, by inverse transform sampling, a cannonical random variable on $D=\realDom$ inducing any given probability distribution on $\realLine$. As usual, we identify the set of real numbers $x\in \realLine$ with the set of maximal elements $\{x\}\in \realDom$. Then $\realLine$ with its Euclidean topology would be a $G_\delta$ subset---i.e., a countable intresection of open subsets---of $\realDom$ equipped with its Scott topology~\cite{Edalat1995}. We say a random variable $r\in (\Omega\to \realDom)$ is {\em supported} on $\realLine$ if $\nu(r^{-1}(\realLine))=1$, which is similar to the corresponding notion for probability distributions~\cite{Edalat1995}. Let $P$ be a probability distribution on $\realLine$ with comulative distribution $F:\realLine\to [0,1]$ given by $F(x)=P((-\infty,x])$, which is a right-continuous increasing function. The inverse distribution (quantile) of $F$ is given by its generalised inverse~\cite{forbes2011statistical}: $G:[0,1]\to \realLine$ with
\[G(p)=\inf \{x:p\leq F(x)\}.\]
 If $F$ is continuous and strictly increasing then $G$ is continuous and is the inverse of $F$. In general, $G$ is right-continuous and increasing, thus has at most a countable set of discontinuities. We have a Galois connection: $G(p)\leq x$ iff $p\leq F(x)$. By Proposition~\ref{envelope}, $G$ has a domain-theoretic extension $G^\star:[0,1]\to \realDom$ which coincides with $G$ wherever $G$ is continuous, and at a point $p$ of disconinutiy of $G$ has the value $G^\star(p)= [x_1,x_2]$ with $x_1=\sup \{x:F(x)<p\}$ and $x_2=\sup \{x:F(x)\leq p\}$. 
\begin{proposition}
The map $G^\star:[0,1]\to \realDom$  is a random variable supported on $\realLine$ with probability distribution $P$. 
\end{proposition}
\subsection{Sampling and normal distribution}
Given any random variable $r:\Omega \to \realDom$, we obtain, by the monad $h$, equivalent but independent random variables $h_1(r)$ and $h_2(r)$.  By iteration, we can obtain $2^n$ independent but equivalent versions of $r$ as $h_{i_n}h_{i_{n-1}}\ldots h_{i_2}h_{i_1}(r)$ for $1\leq i_t\leq 2$ with $1\leq t\leq n$. This scheme can be employed to generate any finite number of independent samples from a random variable, equivalently from a probability distribution.  Using these i.i.d. random variables we can also generate a number of key random variables with given continuous probability distributions.  

In particular, it allows us to obtain a more efficient algorithm for the normal distribution on $\realLine$. Taking $u_1:=h_1\circ \operatorname{Id}=h_1$ and $u_2:=h_2\circ \operatorname{Id}=h_2$, where $\operatorname{Id}:(0,1)\to (0,1)$ is the identity random variable, by the Box-Muller transform we obtain the two independent random variables 
\[z_1=\sqrt{-2\ln u_1}\cos 2\pi u_2\qquad z_2=\sqrt{-2\ln u_2}\cos 2\pi u_1\]
each having a normal distribution. 

Having $n+1$ i.i.d. random variables, $r_i$ for $1\leq i\leq n+1$, each with standard normal distribution, provide us a random variable $r$ with the student t-distribution of degree $n$:
\[r=\frac{r_{n+1}}{\sqrt{n^{-1}\sum_{i=1}^nr_i^2}}\]
Similarly, the chi-squared distribution $r$ with $n$ degrees of freedom can be sampled from the independent random variables $r_i$ with the standard normal distribution:
\[r=\sum_{i=1}^nr_i^2\]
\subsection{Multivariate distributions}
\subsubsection{Multivariate normal distribution} 
Consider the $k$-dimensional multivariate normal distribution $\mathcal{N}({ m},{S})$ with mean ${ m}\in \realLine^k$ and covariance matrix $S\in \realLine^{k\times k}$. Since $S$ is positive semi-definite, by the extended iterative Cholesky's algorithm~\cite{higham1990analysis}, we obtain lower triangular matrix $L\in \realLine^{k\times k}$ with $S=LL^T$. Let $r_i$ for $1\leq i\leq k$ be i.i.d. with the standard normal distribution, generated by the Box-Muller algorithm and the monad $h$, then the random variable $Lr+m$, where $r=(r_1,\ldots,r_k)^T$, has distribution $N(m,S)$. 
\subsubsection{Function of random variables: Dirichlet distribution} 
Let $X=\realLine$ or $X=[0,1]$, and consider a continuous function $f:X^n\to \realDom$ and its maximal (pointwise) extension $\interval f:\interval X^n\to \realDom$. 

Given $k$ real parameters $\alpha_i>0$ and random variables $x_i$ with values in $[0,1]$ for $1\leq i\leq k$ with $k\geq 2$, the Dirichlet distribution~\cite{ng2011dirichlet} is given by 
\[\frac{1}{B(\alpha)}\prod_{1\leq i\leq k}x_i^{\alpha_i-1}\]
where the normalisation constant is expressed in terms of the Gamma function $\Gamma$ by \[B(\alpha)=\frac{\prod_{i=1}^k\Gamma(\alpha_i)}{\Gamma(\sum_{i=1}^k\alpha_i)}\]
The support of the Dirichlet distribution is  on the $(k-1)$-dimensional simplex defined by\[S_k=\{x_i\geq 0:1\leq i\leq k,\;\sum_{i=1}^kx_i=1\}.\] By Corollary~\ref{multi-distribution}, we obtain the domain-theoretic Dirichlet distribution, given for parameter vector $\alpha=(\alpha_1,\ldots,\alpha_k)^T$, by the map 
 \[D_\alpha: (\Omega\to \interval [0,1])^k\to (\Omega\to  \interval [0,1])\]
given by \[D_{\alpha}(r_1,\ldots,r_k)=\lambda\omega.\frac{\prod_{i=1}^k(r_i(\omega))^{\alpha_i-1}}{B(\alpha)}\]
where, for any real number $a\in \realLine$,  the interval, i.e., pointwise extension of the power map $x\mapsto x^a:[0,1]\to \realLine$ is given by $x\mapsto x^a:\interval [0,1]\to \interval [0,1]$ with
\[x^a=\left\{\begin{array}{lll} \left[(x^-)^a, (x^+)^a\right] & & a\geq 0\\[.3ex]
\left[(x^+)^a, (x^-)^a\right]&& a<0\end{array}\right. \]

\section{Further work}\label{further}

An obvious extension of the present work is to introduce a lambda calculus with a probabilistic primitive, i.e. a simple higher-order probabilistic language, and use the PER domains and the $R$ monad to provide a denotational semantics for the new language, together with an operational semantics for the language and an adequacy result for the two semantics; similar work in this area are presented in~\cite{DanosE11,EhrhardPT18,GoubaultJT23,GoubaultV11,HeunenKSY17,HuotLMS23,JiaLMZ21,Vakar2019}. Another follow-up work consists of developing a framework for the conditional probability and Bayesian statistics in this model. Other areas for further research include probabilistic computation on Polish spaces and modelling stochastic processes in this new framework.


\section*{Acknowledgment}
We thank the anonymous reviewers for their helpful feedback.
This work has been partially supported by the Securing Containers (SecCo) project within the SERICS project (PE00000014), under the NRRP MUR programme funded by the EU - NGEU.



\bibliographystyle{ACM-Reference-Format}
 \bibliography{CCCRandomVariables,library_ord}


\end{document}